\documentclass[preprint,aps,prd]{revtex4-1}
\usepackage{graphicx}
\usepackage{slashed}
\usepackage{verbatim} 
\newcommand{\be}{\begin{eqnarray}}
\newcommand{\ee}{\end{eqnarray}}
\usepackage[colorlinks]{hyperref}
\usepackage{cancel,bm} 
\usepackage{amsmath,amssymb}
\usepackage{mathtools}
\usepackage{float}
\usepackage{color}
\usepackage{leftidx}
\usepackage{booktabs}
\usepackage{latexsym}
\usepackage{revsymb}
\usepackage{multirow}
\usepackage{hypcap}
\usepackage{array}
\usepackage{leftidx}
\usepackage{booktabs}
\usepackage{latexsym}
\usepackage{revsymb}
\hypersetup{
    colorlinks=red,
    citecolor=blue,
    filecolor=magenta,      
    urlcolor=red,
}

\begin{document}

\title{  $Cos(2\phi_h)$ asymmetry in $J/\psi$ production  in unpolarized $ep$ collision}

\author{Raj Kishore, Asmita Mukherjee and Mariyah Siddiqah}
\affiliation{ Department of Physics,
Indian Institute of Technology Bombay, Mumbai-400076,
India.}
\date{\today}

\begin{abstract}
We present a calculation of the $cos (2 \phi_h)$ asymmetry in $J/\psi$ production in electron-proton collision at the future electron-ion collider (EIC), a useful channel to probe the gluon TMDs. We calculate the asymmetry at next-to-leading order (NLO) in $\alpha_s$ in the framework of generalized factorization.   The dominating sub-process is $\gamma^* +g \rightarrow J/\psi+g$. The production of $J/\psi$ is calculated in the non-relativistic QCD(NRQCD) framework with the inclusion of both color singlet and color octet contributions.  Numerical estimates of the 
$cos(2\phi_h)$ asymmetry are given in the kinematical region to be accessed by the future EIC. The asymmetry depends on the parameterization  of the gluon TMDs, as well as on the long distance matrix elements (LDMEs). We use  both  Gaussian-type parameterization
and McLerran-Venugopalan model for the TMDs in the kinematical region of small-$x$, where the gluons play a dominant role. We obtain sizable asymmetry, which may be useful to probe the ratio of the linearly polarized  and the unpolarized gluon   distribution in the proton.


\end{abstract}

\maketitle
\raggedbottom 

\section {Introduction}

Quantum Chromodynamics (QCD) is an exceptionally rich and complex theory of strong interactions between quarks and gluons, the fundamental constituents of matter. However, these quarks and gluons do not exist in nature as free particles but are confined inside hadrons, and their fundamental properties can be explored only with the help of scattering processes. Hadron physics community has expanded its inquiry beyond the ordinary one dimensional collinear parton distribution functions (PDFs) in the motion of the parton and its spatial distribution in a direction perpendicular to the momentum of the parent hadron. 

To account for transverse motion, the collinear PDFs were extended to transverse-momentum-dependant PDFs, also referred to as TMDs\cite{ralston1979production,collins1981back,collins1985transverse}. TMDs have attracted an enormous amount of interest and are being  investigated at the current facilities  including JLAB 12 GeV upgrade,
RHIC and are planned to be investigated at the future electron-ion collider (EIC).
They are considered as an extension of the standard, one-dimensional, parton distribution functions (PDFs) to the three-dimensional momentum space. Due to the gauge invariance, the operator definition of TMDs requires the inclusion of gauge links or Wilson lines. Unlike colinear PDFs, which are universal, TMDs are process dependent due to their initial and final state interactions \cite{collins1983relation,boer2000color}, or the gauge links. In other words, the TMDs extracted from semi-inclusive deep inelastic scattering (SIDIS) are  not the same as extracted from Drell-Yan processes due to the difference in gauge links. TMDs are sensitive to the soft gluon exchanges and the color flow in the specific sub-processes in which they are probed. 
SIDIS and Drell-Yan processes provide the majority of experimental data for the extraction of TMDs, where the observables of most interest are the single-spin asymmetries (SSA) and azimuthal  asymmetries, which have been measured and are currently under direct experimental scrutiny \cite{arneodo1987measurement,airapetian2005single,hermes2000evidence}.

A lot of work has recently been done  to extract quark TMDs inside a proton 
from  low  energy  data  from  HERMES, COMPASS  or  JLab  experiments. Contrarily a little is experimentally known about  the gluon TMDs \cite{mulders2001transverse}, as they typically require higher-energy scattering processes and are harder to isolate in comparison to quark TMDs. Each gluon TMD contains multiple gauge links while   
the quark TMDs contain one, due to which  the process dependence of gluon TMDs is more involved than
quark TMDs \cite{buffing2013generalized}. Gluon TMDs parameterize the transverse motion of gluons inside a proton.  In the parton model, the gluon correlator of the unpolarized spin $1/2$ hadron is parameterized in terms of leading
twist transverse momentum dependent (TMD) distribution functions \cite{mulders2001transverse} $i.e.~f_1^g(x,\mathbf{k}_{\perp}^2)$ and $h_1^{\perp g}(x,\mathbf{k}_{\perp}^2)$. The function $f_1^g(x,\mathbf{k}_{\perp}^2)$ represents the probability of finding an unpolarized gluon, within an unpolarized hadron, with a longitudinal momentum fraction $x$ and transverse momentum $k_\perp$, while as $h_1^{\perp g}(x,\mathbf{k}_{\perp}^2)$ represents the distribution of linearly polarized gluons within the unpolarized hadron and is also known as Boer-Mulders function. These are the only two
TMDs of the unpolarized proton that  provide essential knowledge on the transverse dynamics of the gluon content of the proton. They are also important for the proper explanation of gluon-fusion processes at all energies. At small-$x$, it turns out  that the linearly  polarized  distribution  may  reach  its  maximally  allowed  size,  bounded  by  the unpolarized gluon density \cite{mulders2001transverse}. Depending on the gauge links, there are two types of gluon TMDs, Weizs\"acker-Williams (WW)  type where both gauge links are either future or past pointing \cite{Kovchegov:1998bi, McLerran:1998nk}; and dipole type, where one gauge link is future pointing and one 
past \cite{Dominguez:2011br}.   $ep$ collision probes the WW type gluon distribution.  Various methods
have been proposed to measure both $f_1^g(x,\mathbf{k}_{\perp}^2)$ and $h_1^{\perp g}(x,\mathbf{k}_{\perp}^2)$, as well as other gluon TMDs, for example  \cite{boer2012linearly,sun2011gluon,qiu2011probing,zhang2014probing,boer2014impact, boer2014tmd,buffing2013generalized,mukherjee2016probing,echevarria2015qcd,mukherjee2017j,mukherjee2017linearly,boer2017gluon,lansberg2017associated,d2017probing,rajesh2018sivers,bacchetta2020gluon,lansberg2018pinning,kishore2019accessing,sun2013heavy,ma2013transverse,ma2014breakdown,Boer:2016fqd}.  

In the last few years, the distribution of linearly polarised gluons within an unpolarized proton has attracted a lot of interest. Yet they have not been extracted experimentally.   A lot of theoretical investigations has been put forward to probe $h_1^{\perp g}(x,\mathbf{k}_{\perp}^2)$, and a model-independent theoretical upper bound can be found in Ref. \cite{mulders2001transverse, boer2011direct}. They are time-reversal even (T even) object, affecting the unpolarized cross-section of scattering processes, as well as an azimuthal asymmetry of the type $cos(2\phi_h)$ \cite{pisano2013linear}.  Processes like heavy quark pair or dijet
production in SIDIS \cite{pisano2013linear}, diphoton pair \cite{qiu2011probing} and $\Upsilon (1S)$+jet \cite{den2014accessing} production in $pp$ collision have been suggested for extracting $h_1^{\perp g}(x,\mathbf{k}_{\perp}^2)$. It has been seen
in these processes that $h_1^{\perp g}(x,\mathbf{k}_{\perp}^2)$ can be probed 
 by measuring azimuthal asymmetries. It can also be probed in heavy-quark pair production in lepton-proton, proton-proton collisions  \cite{pisano2013linear,efremov2018measure,efremov2018ratio} and associated production of dilepton and $J/\psi$ \cite{lansberg2017associated} as well. Quarkonium production is a useful tool to probe the gluon TMDs  (see \cite{lansberg2020new}for a recent review).  $J/\psi$ production in SIDIS is a good channel, as the effect of the gauge links are  simpler compared to $pp$ collision, and one can assume the generalized factorization. Quite a lot of theoretical work has been done recently in this direction. TMDs can be accessed through this process when the transverse momentum of the produced $J/\psi$ ($P_{h\perp}$) is not very large, $\mid P_{h \perp} \mid  < M$, where $M $ is the mass of $J/\psi$.  Such kinematical region is expected to be accessible at the future electron-ion collider (EIC). In Ref. \cite{mukherjee2017j} the authors probed $h_1^{\perp g}$ in $cos(2\phi_h)$ asymmetry in $J/\psi$ production through the leading order
(LO) process $\gamma^{\ast}+g\rightarrow J/\psi$ at the future EIC at $z=1$, where $z$ is the fraction of photon energy transferred to $J/\psi$. The calculation of $cos(2\phi_h)$ 
asymmetry which directly probes the WW type linearly
polarized gluon distribution at next-to-leading-order(NLO) using color singlet model (CSM) is already calculated in \cite{kishore2019accessing}, and in $J/\psi$ +jet production in $ep$ collision in NRQCD based color octet model in \cite{DAlesio:2019qpk}.   In this work, we extend our study to investigate the $cos(2\phi_h)$ asymmetry in $J/\psi$ production in ep collision within the NRQCD based Color Octet Model(COM) in the kinematical region $z<1$.

The  $J/\psi$ meson is a charm-anti-charm quarkonium bound state with odd charge parity. It has gained a lot of interest, it not only provides knowledge on the strong interactions responsible for hadronization but also acts as a  probe of the perturbative and non-perturbative aspects of QCD. 
The three main models which have been developed to describe the production mechanism of $J/\psi$
meson are, color singlet model (CSM) \cite{berger1981inelastic,baier1981hadronic,baier1982inelastic}, color evaporation model (CEM)\cite{halzen1977cvc,halzen1978hadroproduction} and color octet model(COM)\cite{bodwin1992rigorous}, are all successful but at different energies.
Each of these models has its field of applicability, and they have also achieved considerable success
in describing the experimental data, but which one is the best is still an open problem.  
All these models have one thing in
common, the cross-section  is  factorized  into  a perturbative hard  part  where  the  quarks  and  gluons  form  the  heavy  quark and antiquark pair, with momentum of order M (heavy quark mass) and a  non-perturbative(non-relativistic) soft part where the heavy quark pair forming a bound state at a scale of order $\varLambda _{QCD}$. Former is also known as short-distance part and can be calculated in order $\alpha_s(M)$. All the information of hadronization is encoded in the long-distance matrix elements (LDME), which are usually extracted by fitting experimental data. They describe the transition probability to form the quarkonium state from the heavy quark pair. A particularly elegant approach for separating a perturbative hard  part from the non-perturbative soft part is to recast the analysis in terms of NRQCD \cite{bodwin1995rigorous}, an effective field theory designed precisely for this purpose.

Color Octet Model is based on NRQCD effective field theory. In COM, the heavy quark relative momentum 
$(M_Q\nu)$ is much less than the mass $(M _Q)$ of heavy quark, where $\nu$ is the relative velocity  between the heavy
quark and anti-quark in the centre-of-mass frame of the quarkonium pair. The  
quark potential model calculations indicate that $\nu^2\approx 0.3$ for charmonium.
The cross-section of quarkonium is expressed as an infinite series in the limit $\nu \ll1$. Each term in the series is the product of $Q\bar{Q}$ pair cross-section in a definite state $n=^{2S+1}L_J^{[a]}$ and LDMEs. Here $J, L$ and $S$ are total angular momentum, orbital
angular momentum and spin quantum numbers, respectively, and $a$ is the color multiplicity
carrying a value of $1$ for color singlet and $8$ for color octet state. Both the color singlet and color octet states are included in this model for the  production rate of quarkonium. A good description of $J/\psi$ by the color octet model is given at RHIC energies \cite{cooper2004j}. Azimuthal asymmetries in $J/\psi$ production in these processes may also be used as a tool to get information on the LDMEs in NRQCD \cite{Boer:2021ehu}. 
In our study of $cos(2\phi_h)$ azimuthal asymmetry in $J/\psi$, we will use NRQCD formalism based on color octet model. We will be taking NLO process $\gamma^{\ast}+g\rightarrow J/\psi+g$ into the consideration, and investigate mainly  the small-$x$ region, where the gluon TMDs play a very important role. 

The rest of this paper is organized as follows. The analytical framework of our calculation is discussed in Section II. In Section III, we present our numerical results and finally, in Section IV, we conclude.

\section{Azimuthal asymmetry in $J/\psi$ leptoproduction}
\subsection{Calculation of the cross section and asymmetry}

We consider semi-inclusive production of $J/\psi$ in unpolarized $ep$ collision :
 
\begin{eqnarray}
 e(l)+p(P)\rightarrow e(l')+J/\psi(P_h)+X(P_x)
\end{eqnarray}
where the quantities within the brackets are the four-momentum of corresponding particles and $X$ represents 
the proton remnant. We use the following invariants to describe the kinematics of this
process,
\begin{eqnarray}
 Q^2(=-q^2),~~~W^2=(P+q)^2,~~~z=\frac{P\cdot P_h}{P\cdot q}
\end{eqnarray}
The other two dimensionless invariants are,
\begin{eqnarray}
 y=\frac{P\cdot q}{P\cdot l},~~~~~~x_B=\frac{Q^2}{2P\cdot q},
\end{eqnarray}
The quantity $Q^2$ is  the virtuality of the photon, $W^2$ is virtual photon-target system invariant mass squared, $z$ is the fraction of energy transferred from photon to $J/\psi$ in the frame where the initial proton is at rest, $y$ is the inelasticity variable and has a physical interpretation as the fraction of the energy of the electron transferred to the proton. The variable $x_B(=Q^2/2P\cdot q)$ is the known as Bjoken-$x$.

To study the azimuthal asymmetry, we use a frame  in which the incoming proton and the virtual photon exchanged in the 
process move  in $+z$ and $-z$ directions.
The kinematics here is defined in terms of two light-like vectors with the help of a Sudakov decomposition, here chosen to
be the momentum $P(=n_{-})$ of the incoming proton, and a second vector $n(=n_{+})$, obeying the relations $n\cdot P = 1$ and $n_{+}^2=n_{-}^ 2 = 0$.  Thus one can have the following expressions for the momenta of the incoming proton and virtual photon$(q=l-l')$:
\begin{eqnarray}
 P&=& n_{-}+\frac{M_p^2}{2}n_{+}\approx n_{-}\\
 q&=& -x_Bn_{-}+\frac{Q^2}{2x_{B}}n_{+}\approx-x_BP+(P\cdot q)n_{+}
 \end{eqnarray}
 Where $M_p$ is the mass of proton.
 Moreover, the momenta of incoming and outgoing lepton can be written as:
 \begin{eqnarray}
 l&=&\frac{1-y}{y}x_BP+\frac{1}{y}\frac{Q^2}{2x_B}n+\frac{\sqrt{(1-y)}}{y}Q\hat{l}_{\perp}\nonumber\\
 &=&\frac{1-y}{y}x_BP+\frac{s}{2}n+\frac{\sqrt{(1-y)}}{y}Q\hat{l}_{\perp}\\
 l'&=&\frac{1}{y}x_BP+\frac{1-y}{y}\frac{Q^2}{2x_B}n+\frac{\sqrt{(1-y)}}{y}Q\hat{l}_{\perp}\nonumber\\
 &=&\frac{1}{y}x_BP+(1-y)\frac{s}{2}n+\frac{\sqrt{(1-y)}}{y}Q\hat{l}_{\perp}.
\end{eqnarray}
Here $s=(l+P)^2=2 P\cdot l$ is the squared centre-of-mass energy of the electron-proton scattering
and
$Q^2, s, y$ and $x_B $ are related by the relation $Q^2=s~x_B~y$. 
At leading order, $J/\psi$ is produced by the virtual-photon gluon fusion for which $z=1$ \cite{mukherjee2017j}. 
Since at small-$x$ the proton is rich in gluons, the dominant partonic sub-process at NLO for the $J/\psi$ production is $\gamma^{\ast}(q)+g(k)\rightarrow J/\psi(P_h)+g(p_g)$. In terms of light-like vectors defined above the four momentum of initial state gluon, final state gluon and $J/\psi$ are expressed as:
\begin{eqnarray}
 k&=&xP+k_{\perp}+(k\cdot P-xM_p^2)n\approx xP+k_{\perp}\\
 p_g&=&(1-z)(P\cdot q)n+\frac{\mathbf{p}_{g\perp}^2}{2(1-z)P\cdot q}P+p_{g\perp}\\
 P_h&=&z(P\cdot q)n+\frac{M^2+\mathbf{P}_{h\perp}^2}{2zP\cdot q}P+P_{h\perp}.
\end{eqnarray}
Here $x=k\cdot n$ is the light cone momentum fraction of the gluon, $M$ is the mass of $J/\psi$ and $P_h^2=-\mathbf{P}_{h\perp}^2$. The Mandelstam variables for the partonic sub-process $\gamma^{\ast}(q)+g(k)\rightarrow J/\psi(P_h)+g(p_g)$ become:
\begin{alignat*}{2}
 \hat{s}&=(k+q)^2=q^2+2k\cdot q=\frac{xQ^2}{x_B}-Q^2,\\
 \hat{t}&=(t-P_h)^2=M^2-2k\cdot P_h,\\
 &=M^2-\frac{xzQ^2}{x_B}+2k_{\perp}P_{h\perp}cos(\phi-\phi_h),\\
 \hat{u}&=(q-P_h)^2=M^2+q^2-2q\cdot P_h,\\
 &=M^2-(1-z)Q^2-\frac{M^2+P_{h\perp}^2}{z},
\end{alignat*}
here $\phi$ and $\phi_h$ represent the azimuthal angles of initial gluon and the transverse momentum of  $J/\psi$, respectively.
Due to the presence of the gauge links or initial/final  state interactions, factorization in many of such processes are still not proven. However, $ep$ collision processes are less complicated than $pp$ collisions in terms of color flow.  In \cite{Boer:2020bbd} the process $ep \rightarrow e+J/\psi+X$ is compared with the  SIDIS, for which TMD factorization has been proven. It is argued that the additional scale, namely, the mass of $J/\psi$ does not affect the gauge link structure, and TMD factorization is expected to be valid for this process as well. Following the approach of  Refs. \cite{ boer2012polarized, mukherjee2017j}, we assume the generalized  factorization in the kinematics considered. The differential scattering cross-section can be written as a convolution of leptonic tensor, a soft parton correlator for the incoming hadron and a hard part:
\begin{equation}
\begin{aligned}
 d\sigma={} &\frac{1}{2s}\frac{d^3l'}{(2\pi)^32E_l'}\frac{d^3P_{h}}{(2\pi)^32E_{P_h}}\int \frac{d^3p_g}{(2\pi)^32E_g}
  \int dxd^2\mathbf{k}_{\perp}(2\pi)^4\delta(q+k-P_h-p_g)\\
  &\times \frac{1}{Q^4}\mathcal{L}^{\mu\mu'}(l,q)\Phi^{\nu\nu'}
  (x,\mathbf{k}_{\perp})~\mathcal{M}_{\mu\nu}(\mathcal{M}_{\mu'\nu'})^{\ast}
  \label{ds}
 \end{aligned}
\end{equation}
 The term $\mathcal{M}_{\mu\nu}$ represents  the amplitude of $J/\psi$ production in  the ${\gamma^{\ast}+g\rightarrow J/\psi+g}$ partonic sub-process. The leptonic tensor is given by
\begin{eqnarray}
 \mathcal{L}^{\mu\mu'}(l,q)&=e^2(-g^{\mu\mu'}Q^2+2(l^{\mu}l^{'\mu'}+l^{\mu'}l^{'\mu})),
\end{eqnarray}
where $e$ is the electronic charge. At leading twist, the gluon correlator of the unpolarized proton contains two TMD gluon distribution functions
\begin{eqnarray}
 \Phi_g^{\nu\nu'}(x,\mathbf{k}_{\perp})=-\frac{1}{2x}\bigg\{g_{\perp}^{\nu\nu'}f_1^g(x,\mathbf{k}_{\perp}^2)-\left(\frac{k_{\perp}^{\nu}k_{\perp}^{\nu'}}{M_p^2}+g_{\perp}^{\nu\nu'}\frac{\mathbf{k}_{\perp}^2}{2M_p^2}\right)h_1^{\perp g}(x,\mathbf{k}_{\perp}^2)\bigg\}.
\end{eqnarray}
Here $g_{\perp}^{\nu\nu'}=g^{\nu\nu'}-P^{\nu}n^{\nu'}/P\cdot n-P^{\nu'}n^{\nu}/P\cdot n$. The quantities $f_1^g(x,\mathbf{k}_{\perp}^2) $ and $h_1^{\perp g}(x,\mathbf{k}_{\perp}^2)$ represent the unpolarized and the linearly polarized gluon distribution functions respectively.

\subsection{$J/\psi$ production in NRQCD based color octet model}

\begin{figure}[H]
  \includegraphics[scale=0.58]{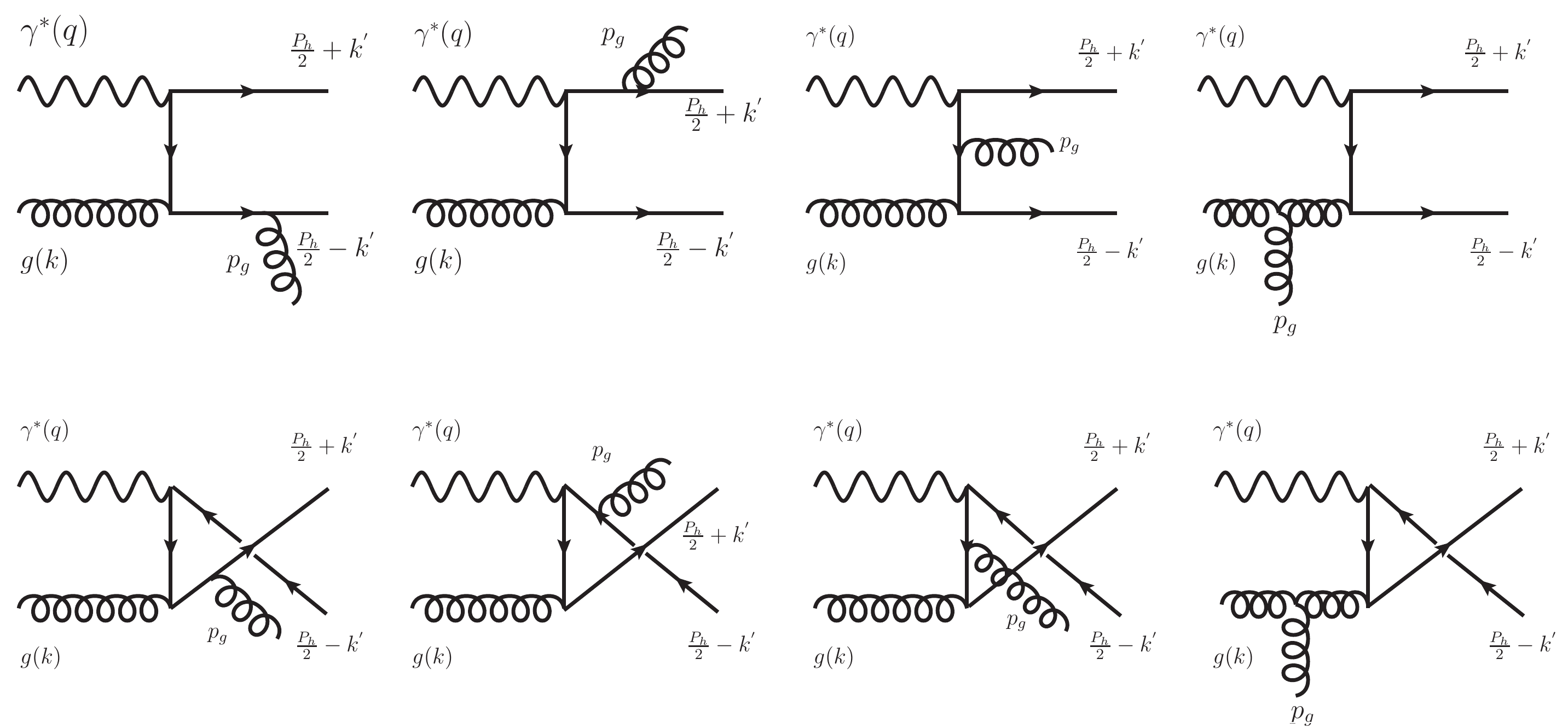}
 \caption{Feynman diagrams for  $\gamma^{\ast}+g\rightarrow J/\psi+g$ at NLO order partonic sub-process.}
 \label{amp}
 \end{figure}

The amplitude for the production of $J/\psi$ can be written as follows  \cite{baier1983hadronic,boer2012polarized} 
\begin{equation}
 \begin{aligned}
  \mathcal{M}\left(\gamma^{\ast}g\rightarrow Q\bar{Q}\left[^{2S+1}L_J^{(1,8)}\right](P_h)+g\right)={}&\sum_{L_zS_z}\int \frac{d^3\mathbf{k}'}{(2\pi)^3}\Psi_{LL_z}(\mathbf{k}')
 \langle LL_z;SS_z|JJ_z\rangle~\\
 &Tr[O(q,k,P_h,k')\mathcal{P}_{SS_z}(P_h,k')].
 \label{13}
 \end{aligned}
\end{equation}
Here $\Psi_{LL_z}(\mathbf{k}')$ is the non-relativistic bound-state wave function with orbital angular momentum $L,L_z$ and the relative momentum $k'$ of the heavy quark in the quarkonium rest frame; $k'$ is assumed to be smaller than $P_h$. 
  $\langle LL_z;SS_z|JJ_z\rangle$ are the usual Clebsch-Gordon coefficients, $O(q,k,P_h,k')$ represents the amplitude for the production of the heavy quark pair  $Q\bar{Q}$ and is calculated from the Feynman diagrams. The Feynman diagrams relevant for this process are given in Fig.~\ref{amp}.  The polarization vectors of the initial gluons and the heavy quark-antiquark legs are absorbed into the
definitions of the gluon correlators and in  the bound state wave function, respectively. The quantity 
\begin{eqnarray}
 \mathcal{P}_{SS_z}(P_h,k')&&=\sum_{s_1,s2}\left\langle\frac{1}{2}s_1;\frac{1}{2}s_2|SS_z\right\rangle v \left(\frac{P_h}{2}-k',s_1\right)\bar{u}\left(\frac{P_h}{2}+k',s_2\right),\nonumber\\
 &&=\frac{1}{4M^{3/2}}(-\slashed{P}_h+2\slashed{k}'+M)\Pi_{SS_z}(\slashed{P}_h+2\slashed{k}'+M)+\mathcal{O}(k^{'2})
\end{eqnarray}
plays the  role of spin projection operator \cite{baier1983hadronic,boer2012polarized}  with
\[
    \Pi_{SS_z}= 
\begin{cases}
    \gamma_5,& \text{for singlet state }(S= 0)\\
    \slashed{\varepsilon}_{S_z}(P_h),              & \text{for triplet state}(S=1),
\end{cases}
\]

where $\varepsilon_{S_z}(P_h)$ is the spin vector of the $Q\bar{Q}$ system. We consider here $\gamma^{\ast}+g\rightarrow J/\psi+g$, which is the dominating  partonic sub-process. 

By considering the contribution from all the above Feynman diagrams we can write the amplitude as
\begin{eqnarray}
 O(q,k,P_h,k')=\sum_{m=1}^8\mathcal{C}_mO_m(q,k,P_h,k'),
\end{eqnarray}
where $\mathcal{C}_m$ represents the color factor of each diagram
\begin{alignat*}{2}
 \mathcal{C}_1=\mathcal{C}_6=\mathcal{C}_7=\sum_{ij}\langle3i;\bar{3}j|8c\rangle(t_at_b)_{ij}\\
 \mathcal{C}_2=\mathcal{C}_3=\mathcal{C}_5=\sum_{ij}\langle3i;\bar{3}j|8c\rangle(t_bt_a)_{ij}\\
 \mathcal{C}_4=\mathcal{C}_8=\sum_{ij}\langle3i;\bar{3}j|8c\rangle if_{abd}(t_d)_{ij},
\end{alignat*}
here the summation runs over the colors of the outgoing quark and antiquark. The $SU(3)$ Clebsch-Gordan coefficients for color singlet and color octet states, respectively, are given as
\begin{alignat*}{2}
 \langle3i;\bar{3}j|1\rangle=\frac{\delta_{ij}}{\sqrt{N_c}},~~~\langle3i;\bar{3}j|8a\rangle=\sqrt{2}(t^a)^{ij}
\end{alignat*}
and they project out the color state of the $Q\bar{Q}$ pair either in the color singlet or in the color octet state. Here, $N_c$ represents the number of colors. In the fundamental representation the generators of the $SU(3)$ group is denoted by $t_a$ following which $Tr(t_at_b)=\delta_{ab}/2$ and $Tr(t_at_bt_c)=1/4(d_{abc}+if_{abc})$. In case of color octet state, the color factors for the production of initial $Q\bar{Q}$ is
\begin{alignat*}{2}
 \mathcal{C}_1=\mathcal{C}_6=\mathcal{C}_7=\frac{\sqrt{2}}{4}(d_{abc}+if_{abc}),\\
 \mathcal{C}_2=\mathcal{C}_3=\mathcal{C}_5=\frac{\sqrt{2}}{4}(d_{abc}-if_{abc}),\\
 \mathcal{C}_4=\mathcal{C}_8=\frac{\sqrt{2}}{2}if_{abc}.
 \end{alignat*}
The amplitudes $O_m(q,k,P_h,k')$ for the above Feynman diagrams are written as:
\begin{eqnarray}
 O_1&&=4g_s^2(ee_c)\varepsilon_{\lambda_a}^{\mu}(k)\varepsilon_{\lambda_b}^{\nu}(q)\varepsilon_{\lambda_g}^{\rho\ast}(p_g)\gamma_{\nu}\frac{\slashed{P}_h+2\slashed{k}'-2\slashed{q}+M}{(P_h+2k'-2q)^2-M^2}\gamma_{\mu}
 \frac{-\slashed{P}_h+2\slashed{k}'-2\slashed{p}_g+M}{(P_h-2k'+2p_g)^2-M^2}\gamma_{\rho},\\
  O_2&&=4g_s^2(ee_c)\varepsilon_{\lambda_a}^{\mu}(k)\varepsilon_{\lambda_b}^{\nu}(q)\varepsilon_{\lambda_g}^{\rho\ast}(p_g)\gamma_{\rho}
\frac{\slashed{P}_h+2\slashed{k}'+2\slashed{p}_g+M}{(P_h+2k'+2p_g)^2-M^2}
\gamma_{\nu}
  \frac{-\slashed{P}_h+2\slashed{k}'+2\slashed{k}+M}{(P_h-2k'-2k)^2-M^2}
\gamma_{\mu},\\
  O_3&&=4g_s^2(ee_c)\varepsilon_{\lambda_a}^{\mu}(k)\varepsilon_{\lambda_b}^{\nu}(q)\varepsilon_{\lambda_g}^{\rho\ast}(p_g)\gamma_{\nu}\frac{\slashed{P}_h+2\slashed{k}'-2\slashed{q}+M}{(P_h+2k'-2q)^2-M^2}\gamma_{\rho}
 \frac{-\slashed{P}_h+2\slashed{k}'+2\slashed{k}+M}{(P_h-2k'-2k)^2-M^2}\gamma_{\mu},\\
 O_4&&=2g_s^2(ee_c)\varepsilon_{\lambda_a}^{\mu}(k)\varepsilon_{\lambda_b}^{\nu}(q)\varepsilon_{\lambda_g}^{\rho\ast}(p_g)\gamma_{\nu}\frac{\slashed{P}_h+2\slashed{k}'-2\slashed{q}+M}{(P_h+2k'-2q)^2-M^2}\gamma_{\sigma}
  \frac{1}{(k-p_g)^2}\mathcal{T}_{\mu\rho\sigma}(k,p_g).
\end{eqnarray}
 Here $M=2m_c$, with $m_c$ representing the charm quark mass and the tensor $\mathcal{T}_{\mu\rho\sigma}(k,p_g)=g_{\mu\rho}(k+p_g)_{\sigma}+g_{\rho\sigma}(k-2p_g)_{\mu}+g_{\sigma\mu}(p_g-2k)_{\rho}$ is the three gluon vertex. As all the Feynman diagrams are symmetric so we can obtain the rest of four amplitudes $O_5,O_6,O_7$ and $O_8$ by reversing the fermion flow and replacing $k'$ by $-k'$. 
 In the rest frame of the bound
state, the relative momentum of $k'$ is very small as compared to $P_h$, thus Taylor expansion can be performed in powers of $k'$ around $k'=0$ in Eq. \ref{13}. So it becomes convenient to split the radial and angular parts of Fourier transformation of the wave function $\Psi_{LL_z}(\mathbf{k}')$,
\begin{alignat*}{2}
 \int\frac{d^3\mathbf{k}'}{(2\pi)^3}e^{i\mathbf{k}'.\mathbf{r}}\Psi_{LL_z}(\mathbf{k}')=\tilde{\Psi}_{LL_z}(\mathbf{r})=R_L(|r|)Y_{LL_z}(\theta,\varphi),
\end{alignat*}
where $\mathbf{r}=(|\mathbf{r}|,\theta,\varphi)$ in spherical coordinates, $R_L(|r|)$ and 
$Y_{LL_z}(\theta,\varphi)$ is the radial wave function and spherical harmonic respectively. In particular this means,
\begin{eqnarray}
 \int \frac{d^3\mathbf{k'}}{(2\pi)^3}\Psi_{00}(\mathbf{k}')=\frac{1}{\sqrt{4\pi}}R_0(0)
\end{eqnarray}
As the first term of Taylor expansion at $\mathbf{k}'=0$ does not depend on $\mathbf{k}'$ anymore, so it gives the contribution from $S$ waves $(L=0,J=0,1)$,
\begin{eqnarray}
 \mathcal{M}\left[^{2S+1}S_{J}^{(8)}\right](P_h,k)&&=\frac{1}{\sqrt{4\pi}}R_0(0)Tr\left[O(q,k,P_h,k')\mathcal{P}_{SS_z}(P_h,k')\right]\lvert_{k'=0}\nonumber\\
 &&=\frac{1}{\sqrt{4\pi}}R_0(0)Tr\left[O(0)\mathcal{P}_{SS_z}(0)\right],
\end{eqnarray}
where $O(0)=(q,k,P_h,0)$ and $\mathcal{P}_{SS_z}(0)=\mathcal{P}_{SS_z}(P_h,0)$. For $P$ waves $(L=1,J=0,1,2)$, $R_1(0)=0$, so in the Taylor expansion the terms linear in $k^{\alpha}$ are considered in Eq. \ref{13}. Thus,
\begin{eqnarray}
 \int \frac{d^3\mathbf{k}'}{(2\pi)^3}k^{\alpha}\Psi_{1L_z}(\mathbf{k})=-i\varepsilon_{L_z}^{\alpha}(P_h)\sqrt{\frac{3}{4\pi}}R'_1(0)
\end{eqnarray}
where $\varepsilon_{L_z}^{\alpha}(P_h)$ is a polarisation vector for $L=1$ bound state and $R'_1(0)$ represents the derivative of $P$-wave (radial) wave function evaluated at origin. Thus
\begin{eqnarray}
 \mathcal{M}\left[^{2S+1}P_J^{(8)}\right]&&=-i\sqrt{\frac{3}{4\pi}}R'_1(0)
  \sum_{L_zS_z}\varepsilon^{\alpha}_{L_z}(P_h)
\langle LL_z;SS_z\lvert JJ_z\rangle\frac{\partial}{\partial k^{'\alpha}}Tr[O(q,k,P_h,k')\mathcal{P}_{SS_z}(P_h,k')]\lvert_{k'=0}\nonumber\\
  &&=-i\sqrt{\frac{3}{4\pi}}R'_1(0)\sum_{L_zS_z}\varepsilon^{\alpha}_{L_z}(P_h)\langle LL_z;SS_z\lvert JJ_z\rangle Tr[O_{\alpha}(0)\mathcal{P}_{SS_z}(0)+O(0)\mathcal{P}_{SS_z\alpha}(0)]
\end{eqnarray}
where
\begin{eqnarray}
 O_{\alpha}(0)&&=\frac{\partial}{\partial k^{'\alpha}}O(q,k,P_h,k')\lvert_{k'=0},\nonumber\\
 \mathcal{P}_{SS_z\alpha}(0)&&=\frac{\partial}{\partial k^{'\alpha}}\mathcal{P}_{SS_z}(P_h,k')\lvert_{k'=0}\nonumber
\end{eqnarray}
For $P$-wave we could utilize the following relations for Clebsch-Gordan coefficients and various polarisation vectors \cite{kuhn1979electromagnetic,guberina1980rare}:
\begin{eqnarray}
 \sum_{L_zS_z} \langle 1L_z;SS_z|00\rangle \varepsilon^\alpha_{ s_z }(P_h)\varepsilon^\beta_{ L_z }(P_h)&&
=\sqrt{\frac13}\left(g^{\alpha\beta}-\frac{1}{M^2}P_h^\alpha P_h^\beta\right),\\
\sum_{L_zS_z} \langle 1L_z;1S_z|1J_z\rangle \varepsilon^\alpha_{ s_z }(P_h)\varepsilon^\beta_{ L_z }(P_h)&&
=-\frac{i}{M}\sqrt{\frac12}\epsilon_{\delta\lambda\rho\sigma}g^{\rho\alpha}g^{\sigma\beta}
P_h^\delta\varepsilon^\lambda_{ J_z }(P_h),\\
\sum_{L_zS_z} \langle 1L_z;1S_z|2J_z\rangle \varepsilon^\alpha_{ s_z }(P_h)\varepsilon^\beta_{ L_z }(P_h)&&
=\varepsilon^{\alpha\beta}_{ J_z }(P_h).
\end{eqnarray}
Here the term $\varepsilon^\alpha_{ J_z }(P_h)$ represents the polarization vector of bound state with $J=1$, obeying the following relations:
\begin{eqnarray}
 \varepsilon^\alpha_{ J_z }(P_h)P_{h\alpha}&=&0, \nonumber\\
\sum_{L_z}\varepsilon^\alpha_{ J_z }(P_h)\varepsilon^{\ast\beta}_{ J_z }(P_h)&=&
-g^{\alpha\beta}+\frac{P_h^\alpha P_h^\beta}{M^2}\equiv\mathcal{Q}^{\alpha\beta}.
\end{eqnarray}
while as $\varepsilon^{\alpha\beta}_{ J_z }(P_h)$ is the polarization tensor for $J=2$ bound state and 
obeys the following set of relations \cite{kuhn1979electromagnetic,guberina1980rare}:
\begin{eqnarray}
\varepsilon^{\alpha\beta}_{ J_z }(P_h)=\varepsilon^{\beta\alpha}_{ J_z }(P_h), ~~~~
\varepsilon^{\alpha}_{ J_z\alpha }(P_h)=0,~~~
P_{h\alpha}\varepsilon^{\alpha}_{ J_z}(P_h)=0,~~~\nonumber\\
\varepsilon^{\mu\nu}_{ J_z }(P_h)\varepsilon^{\ast\alpha\beta}_{ J_z }(P_h)=\frac12[\mathcal{Q}^{\mu\alpha}
\mathcal{Q}^{\nu\beta}+\mathcal{Q}^{\mu\beta}\mathcal{Q}^{\nu\alpha}]-\frac13\mathcal{Q}^{\mu\nu}
\mathcal{Q}^{\alpha\beta}.
\end{eqnarray}
The relation between the radial wave function $R_0(0)$ and its derivative at origin $R'_1(0)$ with the LDME  can be found in the Eqs.$(36-38)$ within Ref. \cite{rajesh2018sivers}. 

The numerical values of LDMEs for these color octet states are extracted from Ref.\cite{chao2012j}. Now using the above formalism together with the symmetry relations and sum over the color factors from Ref.\cite{rajesh2018sivers}  in the Eqs.$ (39,40,43,44,47)$, we could write the amplitude for color octet states $(^3S_1,^1S_0,^3P_{J(0,1,2)})$. 
\subsubsection{$^3S_1$ Amplitude}
The final expression for the ${^3}{S}{_1}$ scattering amplitude:
\begin{eqnarray}\label{e16}
\begin{aligned}
 \mathcal{M}[\leftidx{^{3}}{S}{_1}^{(8)}](P_h,k)=&{}\frac{1}{4\sqrt{\pi M}}R_0(0)\frac{\sqrt{2}}{2}d_{abc}
 \mathrm{Tr}\left[\sum_{m=1}^3O_m(0)(-\slashed{P}_h+M)\slashed{\varepsilon}_{s_z}\right],
\end{aligned}
\end{eqnarray}

where
\begin{eqnarray}\label{e17}
\sum_{m=1}^3O_m(0)&=&{}g^2_s(ee_c)\varepsilon^\mu_{\lambda_a}(k)\varepsilon^\nu_{\lambda_b}(q)\varepsilon^{
\rho\ast}_{
\lambda_g}(p_g)\Bigg[\frac{\gamma_\nu(\slashed{P_h}-2\slashed{q}+M)\gamma_\mu
(-\slashed{P_h}-2\slashed{p}_g+M)\gamma_\rho}{(\hat{s}-M^2)(\hat{u}-M^2)}\nonumber\\
&+&\frac{\gamma_\rho(\slashed{P_h}+2\slashed{p}_g+M)\gamma_\nu
(-\slashed{P_h}+2\slashed{k}+M)\gamma_\mu}{(\hat{s}-M^2)(\hat{t}-M^2)}\nonumber\\ &+&
\frac{\gamma_\nu(\slashed{P_h}-2\slashed{q}+M)\gamma_\rho(-\slashed{P_h}+2\slashed{k}+M)\gamma_\mu
}{(\hat{t}-M^2)(\hat{u}-M^2)}
\Bigg].
\end{eqnarray}
 Due to the symmetry relations  \cite{rajesh2018sivers} Feynman diagrams $4$ and $8$ cancel and  do not contribute to ${^3}{S}{_1}$ state.
\subsubsection{  ${^1}{S}{_0}$ Amplitude}
For the ${^1}{S}{_0}$  scattering amplitude the expression reads as:
\begin{eqnarray}\label{e21}
\begin{aligned}
 \mathcal{M}[\leftidx{^{1}}{S}{_0}^{(8)}](P_h,k)=&{}\frac{R_0(0)}{4\sqrt{\pi M}}\frac{\sqrt{2}}{2}if_{abc}
 \mathrm{Tr}\big[\left(O_1(0)-O_2(0)-O_3(0)+2O_4(0)\right)
 (-\slashed{P}_h+M)\gamma^5\big]
\end{aligned}
\end{eqnarray}
where $O_1(0),O_2(0)$ and $O_3(0)$ are given in Eq.\eqref{e17} and 
\begin{equation}\label{ee4}
 \begin{aligned}
O_4(0)= 
g^2_s(ee_c)\varepsilon^\mu_{\lambda_a}(k)\varepsilon^\nu_{\lambda_b}(q)\varepsilon^{\rho\ast}_{
\lambda_g}(p_g)
\frac{\gamma_\nu(\slashed{P_h}-2\slashed{q}+M)\gamma^\sigma}{\hat{u}(\hat{u}-M^2)}
\mathcal{T}_{\mu\rho\sigma}(k,p_g)
\end{aligned}
\end{equation}

\subsubsection{  ${^3}{P}{_J}$ Amplitude}
The amplitude for $^3P_J$ can be written as,
\begin{eqnarray}\label{3pj8} 
\mathcal{M}[\leftidx{^{3}}{P}{_J}^{(8)}](P_h,k)&=&\frac{\sqrt{2}}{2}f_{
abc }\sqrt{\frac{3}{4\pi}}R^\prime_1(0)\sum_{L_zS_z}
\varepsilon^\alpha_{L_z}(P_h)
\langle 1L_z;1S_z|JJ_z\rangle\nonumber\\ & &
\mathrm{Tr}\bigg[\left(O_{1\alpha}(0)-O_{2\alpha}(0)-O_{3\alpha}(0)+2O_{4\alpha}(0)\right)\mathcal{P}_{SS_z}
(0)\nonumber\\
&+&\left( O_1(0)-O_2(0)-O_3(0)+2O_4(0)\right)
\mathcal{P}_{SS_z\alpha}(0)
 \bigg].
\end{eqnarray}
 
\subsection{$cos(2\phi_h)$ Azimuthal Asymmetry }

To probe the TMDs in this process, one needs two well-separated scales. While $Q^2$ gives the hard scale, the other scale 
is given by $P_{h \perp}$, and we consider a kinematical region where the transverse momentum  of $J/\psi$ is small compared 
to the mass of $J/\psi$, $M$ $,~ i.e. ~~ P_{h\perp}<M$.  For large transverse momentum of $J/\psi$ collinear factorization is expected to hold.  The generic structure of the TMD cross-section defined by Eq. \ref{ds} is easily obtained from the contraction of four tensors which are
\begin{eqnarray}
L^{\mu\mu'}(l,q)\Phi^{\nu\nu'}(x,{\bm k}_{\perp})\mathcal{M}^{\gamma^*+g\rightarrow J/\psi +g}_{\mu\nu}\mathcal{M}^{*\gamma^*+g\rightarrow J/\psi +g}_{\mu'\nu'}
\end{eqnarray}
We get a contribution for the amplitude and their corresponding complex conjugate amplitude from all the six states ${^1}S_0^{(8)},^3S_1^{(1,8)}$ and $^3P_{J(=0,1,2)}^{(8)}$ in the cross section.  The summation over the transverse polarization of the final on-shell gluon is given by
\begin{eqnarray}
\sum_{\lambda_a=1}^2\varepsilon^{\lambda_a}_\mu(p_g)\varepsilon^{\ast\lambda_a}_{\mu^\prime}(p_g)=-g_{
	\mu\mu^\prime}+
\frac{p_{g\mu} n_{g\mu^\prime}+p_{g\mu^\prime}n_{g\mu}}{p_g\cdot n_g}-\frac{p_{g\mu} p_{g\mu^\prime}}{(p_g\cdot n_g)^2}
\end{eqnarray}
with $n_g^\mu=\frac{P^\mu_h}{M}$.
In a frame where the virtual photon and target proton move along the $z$-axis, and the lepton scattering plane defines the azimuthal angles $\phi_l=\phi_{l'}=0$, we can write: 
\begin{eqnarray}
\frac{d^3l'}{(2\pi)^32E_{l'}}&&=\frac{1}{16\pi^2}sydx_Bdy,~\frac{d^3P_h}{(2\pi)^32E_{h}}=\frac{1}{(2\pi)^3}\frac{1}{2z}dzd^2\textbf{P}_{h\perp},~\nonumber\\&&
 \frac{d^3p_g}{(2\pi)^32E_{g}}=\frac{1}{(2\pi)^3}\frac{1}{2z_2}dz_2d^2\textbf{p}_{g\perp} 
\end{eqnarray}
and the delta function can be written as
\begin{eqnarray}
\begin{split}
\delta^4(q+k-P_h-p_g)=&\delta\Big ( x-\frac{1}{ys}(x_Bys+\frac{M^2+P_{h\perp}^2}{z}+\frac{(k_{\perp}-P_{h\perp})^2}{(1-z)}) \Big )\\
&\times\frac{2}{ys}\delta(1-z-z_2) \times\delta^2(\textbf{k}_{\perp}-\textbf{P}_{h \perp}-\textbf{p}_{g\perp})
\end{split}
\end{eqnarray}
here, the delta function sets $z_2=(1-z)$. After integration over  $x$, $z_2$ and $p_{g\perp}$, 
the final form of the differential cross section can be written as
\begin{eqnarray}
\frac{d\sigma}{dydx_Bdzd^2\textbf{P}_{h\perp}}=\frac{1}{256\pi^4}\frac{1}{x_B^2s^3y^2z(1-z)}\int k_{\perp}dk_{\perp}
{\mid M' \mid }^2
\end{eqnarray}
where, ${\mid M' \mid }^2=\int d\phi{\mid M \mid }^2$, and $k_\perp$ is the
magnitude of ${\bf k_\perp}$. We only keep the transverse momentum of initial gluon up to $\mathcal{O}(k_{\perp}^2/M_p^2)$. We expand in $P_{h\perp}/M$ and  keep the terms up to $\mathcal{O}(P^2_{h\perp}/M^2)$. As we are interested in the small-$x$ domain, we have expanded the amplitudes in $x_B$ and did not consider higher order terms in $x_B$. 
 Thus we could write the final expression for differential scattering cross-section as
\begin{eqnarray} \label{csf1}
\frac{d\sigma}{dydx_Bdzd^2\textbf{P}_{hT}}=d\sigma^{U}(\phi_h)+d\sigma^{T}(\phi_h),
\end{eqnarray}
where 
\begin{eqnarray} 
d\sigma^{U}(\phi_h)=\frac{1}{256\pi^4}\frac{1}{x_B^2s^3y^2z(1-z)}\int k_{\perp}dk_{\perp}
\{(A_0+A_1cos(\phi_h)+A_2cos(2\phi_h))f_1^g(x,\textbf{k}_{\perp}^2) \}\nonumber
\end{eqnarray}
and 
\begin{eqnarray}
d\sigma^{T}(\phi_h)=\frac{1}{256\pi^4}\frac{1}{x_B^2s^3y^2z(1-z)}\int dk_{\perp}\frac{k_{\perp}^3}{M_p^2}\{(B_0+B_1cos(\phi_h)+B_2cos(2\phi_h))h_1^{\perp g}(x,\textbf{k}_{\perp}^2) \}.\nonumber
\end{eqnarray}

 The  analytic expressions of the coefficients $A_0, A_1,A_2, B_0, B_1$ and $B_2$ are very large, which we have not included in this article.  They are available upon request. The $cos(2\phi_h)$ asymmetry measured in experiments is defined as 
\begin{eqnarray}
\langle cos(2\phi_h)\rangle&=&\frac{\int d\phi_h cos(2\phi_h)d\sigma}{\int d\phi_h d\sigma},
\end{eqnarray}
where $\phi_h$ is the azimuthal angle of $J/\psi$ production plane with the lepton plane. The $cos(2\phi_h)$ asymmetry as a function of $P_{h\perp},x_B,z$ and $y$ can be written as:
\begin{eqnarray}
 \langle cos(2\phi_h)\rangle\propto\frac{ \int k_\perp d k_\perp \Big (  A_2~f_1^g(x,\textbf{k}_{\perp}^2)+\frac{k_{\perp}^2}{M_p^2}~B_2~h_1^{\perp g}(x,\textbf{k}_{\perp}^2) \Big ) }{\int k_\perp d k_\perp \Big (A_0~f_1^g(x,\textbf{k}_{\perp}^2)+\frac{k_{\perp}^2}{M_p^2}~B_0~h_1^{\perp g}(x,\textbf{k}_{\perp}^2) \Big )}.
\end{eqnarray}
Thus when the color octet contributions are taken into account  the unpolarized gluon distribution and the linearly polarized distributions are not disentangled in this process in the kinematics considered. However, the above asymmetry depends on the ratio of the two, and can be used to extract this ratio.

\section{ Results and discussion}

In this section, we present numerical estimates of the $cos(2\phi_h)$ asymmetry in the kinematical region to be accessed at EIC. 
As gluon TMDs are particularly important in the small-$x$ domain, we are considering the small-$x$ kinematics. At $z=1$, one gets contribution from the leading order as well as diffractive contributions.  Here the momentum fraction of the final gluon is $(1-z)$. This means as $z \rightarrow1$, the final gluon becomes soft. To keep the final gluon hard, we have imposed a cutoff $z<0.9$. Also, to avoid the gluon fragmentation contribution to $J/\psi$, we have used a lower cutoff $z>0.1$. We have checked that changing the lower cutoff  does not affect the asymmetry much. We took mass of the proton to be $M_p = 1$ GeV.   

The contraction in the calculation above for the different states  $i.e.,~{^1}S_0^{(8)}, ^3S_1^{(1,8)}$ and $^3P_{J(=0,1,2)}^{(8)}$ is calculated using the FeynCalc \cite{shtabovenko2020feyncalc,mertig1991feyn}. In all the plots of the asymmetry, the  
long distance matrix elements (LDMEs) are taken from Ref. \cite{chao2012j} except for the right panel of Fig.~\ref{g2} and for the plots in Fig.~\ref{g4}, where we have used different sets of LDMEs. 
The asymmetry depends on the parameterization/model used for the gluon TMDs. In our estimate, we have used  two sets of parameterization to calculate the $cos(2\phi_h)$ asymmetry:~1) Gaussian-type parameterization \cite{boer2012polarized,mukherjee2017linearly,mukherjee2016probing} for  both the unpolarized as well as the linearly polarized 
TMDs and~ 2) McLerran-Venugopalan model \cite{mclerran1994computing,mclerran1994gluon,mclerran1994green} at small-$x$ region.  
\subsubsection{$cos(2\phi_h)$ asymmetry in the Gaussian parameterization}

\begin{figure}[H]
 
 \centering
 \includegraphics[width=.50\textwidth]{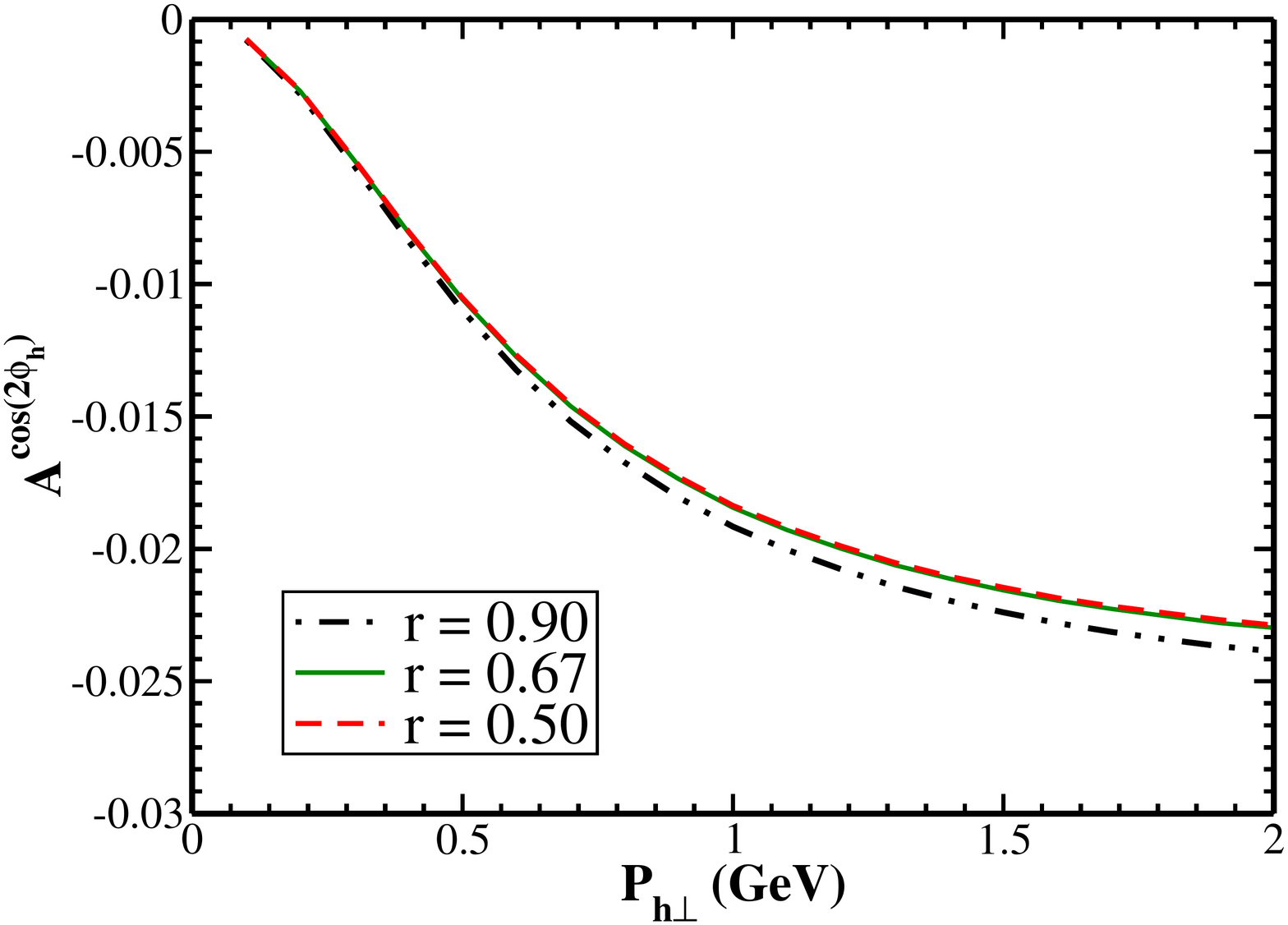}\hfill
    \includegraphics[width=.50\textwidth]{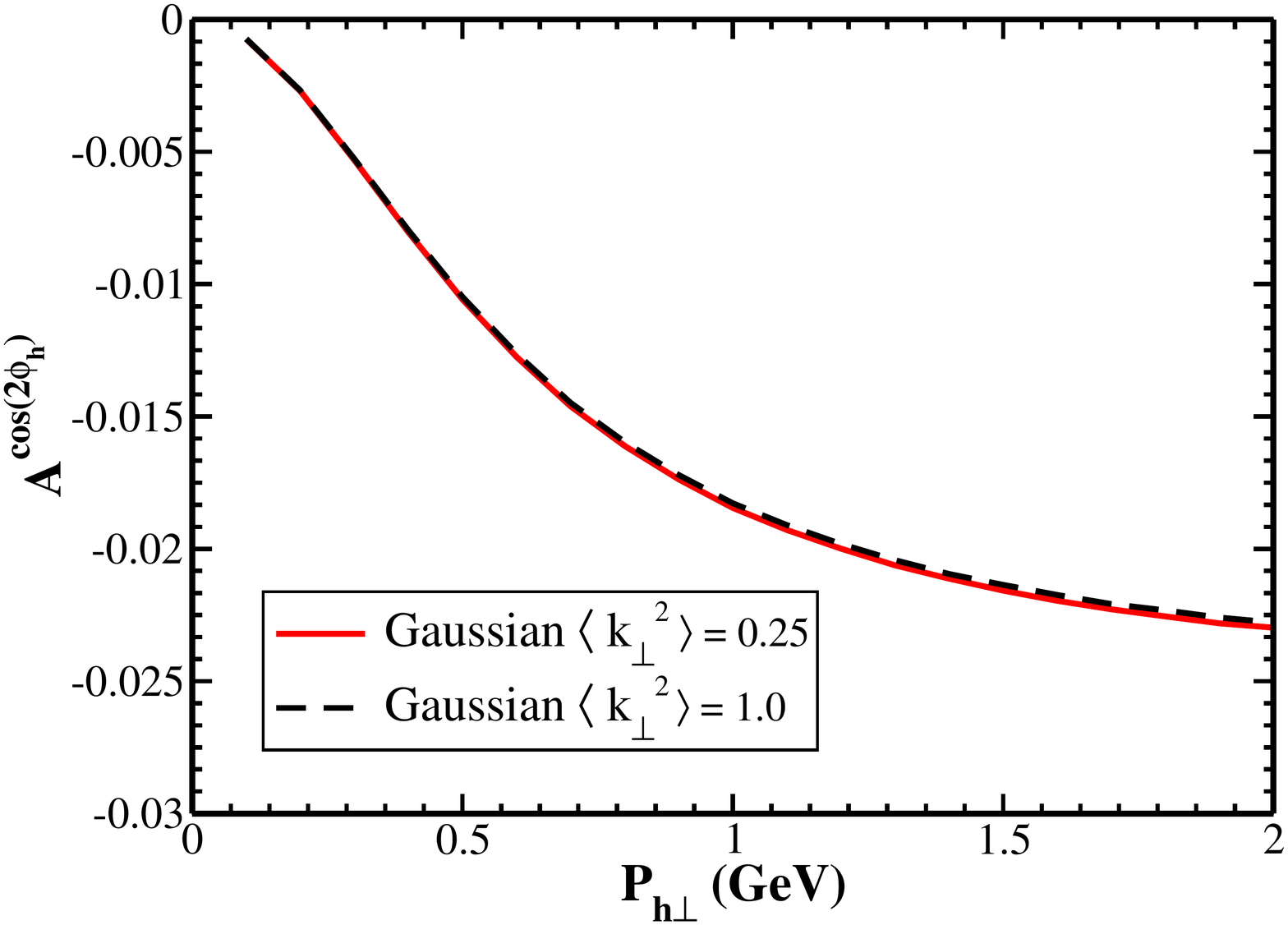}\hfill

 \caption{$cos(2\phi_h)$ asymmetry  in $e+p\rightarrow e+ J/\psi +X$
		process as function of $P_{h\perp}$ at $\sqrt{s}=100$ GeV and $Q^2=15$ GeV$^2$.
 		The integration ranges are $0.0015 < x_B < 0.1$, $0.1 < z < 0.9$ and y is set by the values of $Q^2$, $s$ and $x_B$.
  		Left:  asymmetry for three different values of $r$. Right: 
  		asymmetry for two different values of Gaussian width parameter. For both plots CMSWZ set of LDMEs \cite{chao2012j} is used.}
 \label{g3}
 \end{figure}

In Gaussian type of parameterization, both the TMDs, $f_1^g$ and $h_1^{\perp g}$ are factorized into a product of collinear PDFs times exponential factor which is a function of the transverse momentum.
\begin{eqnarray}
\label{g1}
 f_1^g(x,\textbf{k}_{\perp}^2)=f_1^g(x,\mu)\frac{1}{\pi\langle k_{\perp}^2\rangle}e^{-k_{\perp}^2/\langle k_{\perp}^2\rangle}
 \end{eqnarray}
 \begin{eqnarray}
 \label{g2}
h_1^{\perp g}(x,\textbf{k}_{\perp}^2)=\frac{M_p^2f_1^g(x,\mu)}{\pi\langle k_{\perp}^2\rangle^2}\frac{2(1-r)}{r}e^{1-\frac{k_{\perp}^2}{r\langle k_{\perp}^2\rangle}},
\end{eqnarray}
 $r(0<r<1)$ is a parameter. We took the Gaussian width ${\langle k_{\perp}^2\rangle}=0.25~ \mathrm{GeV}^2$ and $r=1/3$ in all plots except Fig.~\ref{g3}. The term  $f_1^g(x,\mu)$ is the collinear PDF, for which MSTW2008 \cite{martin2009parton} is used and is probed at the scale $\mu=\sqrt{M^2+P_{h\perp}^2}$, where $M(=3.096 $ GeV) is the mass of $J/\psi$. The linearly polarized gluon distribution in the parameterization above satisfies the positivity bound \cite{mulders2001transverse}, but does not saturate it. 

\begin{figure}[H]
 
 \centering
 \includegraphics[width=.5\textwidth]{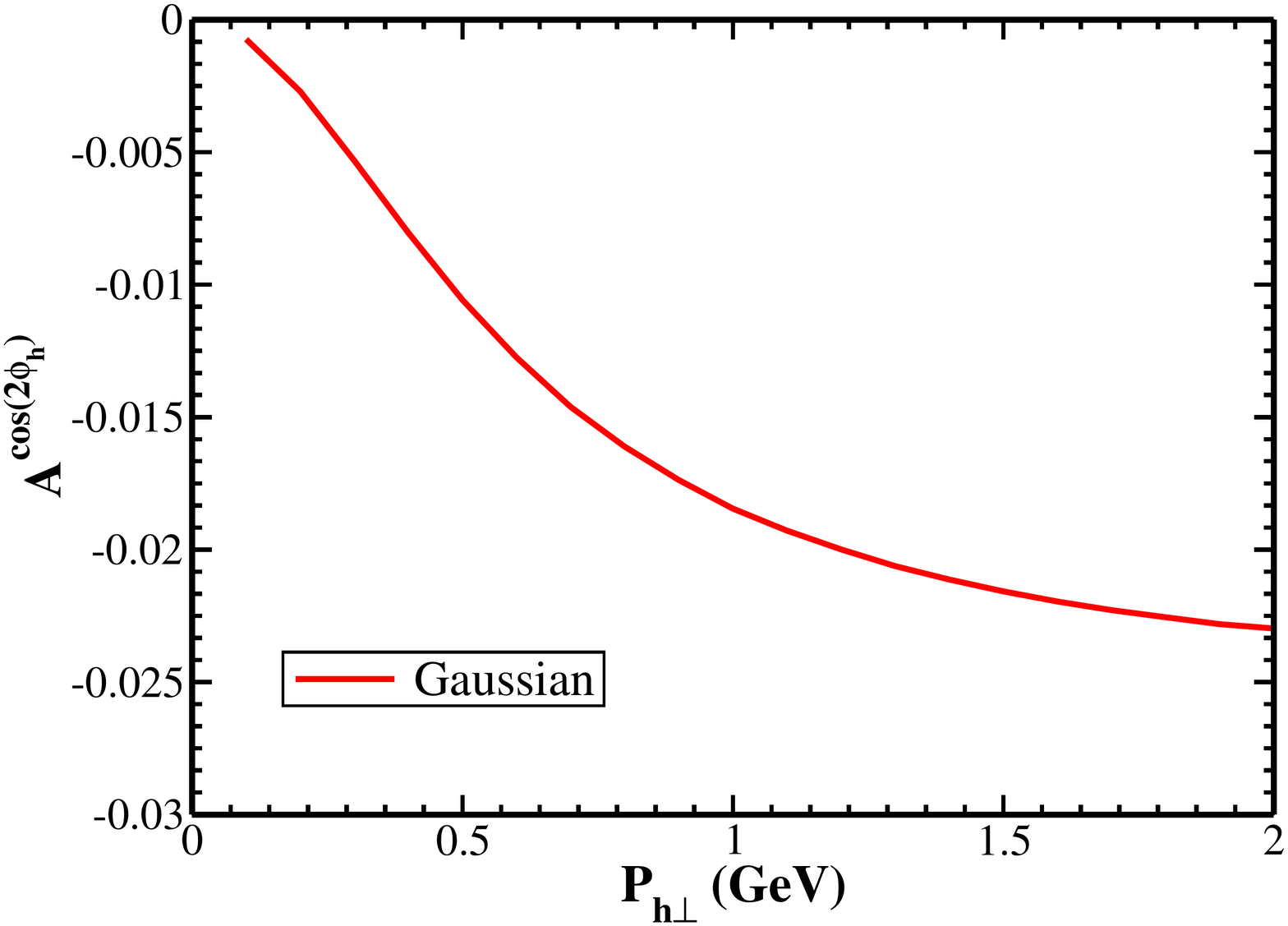}\hfill
  \includegraphics[width=.5\textwidth]{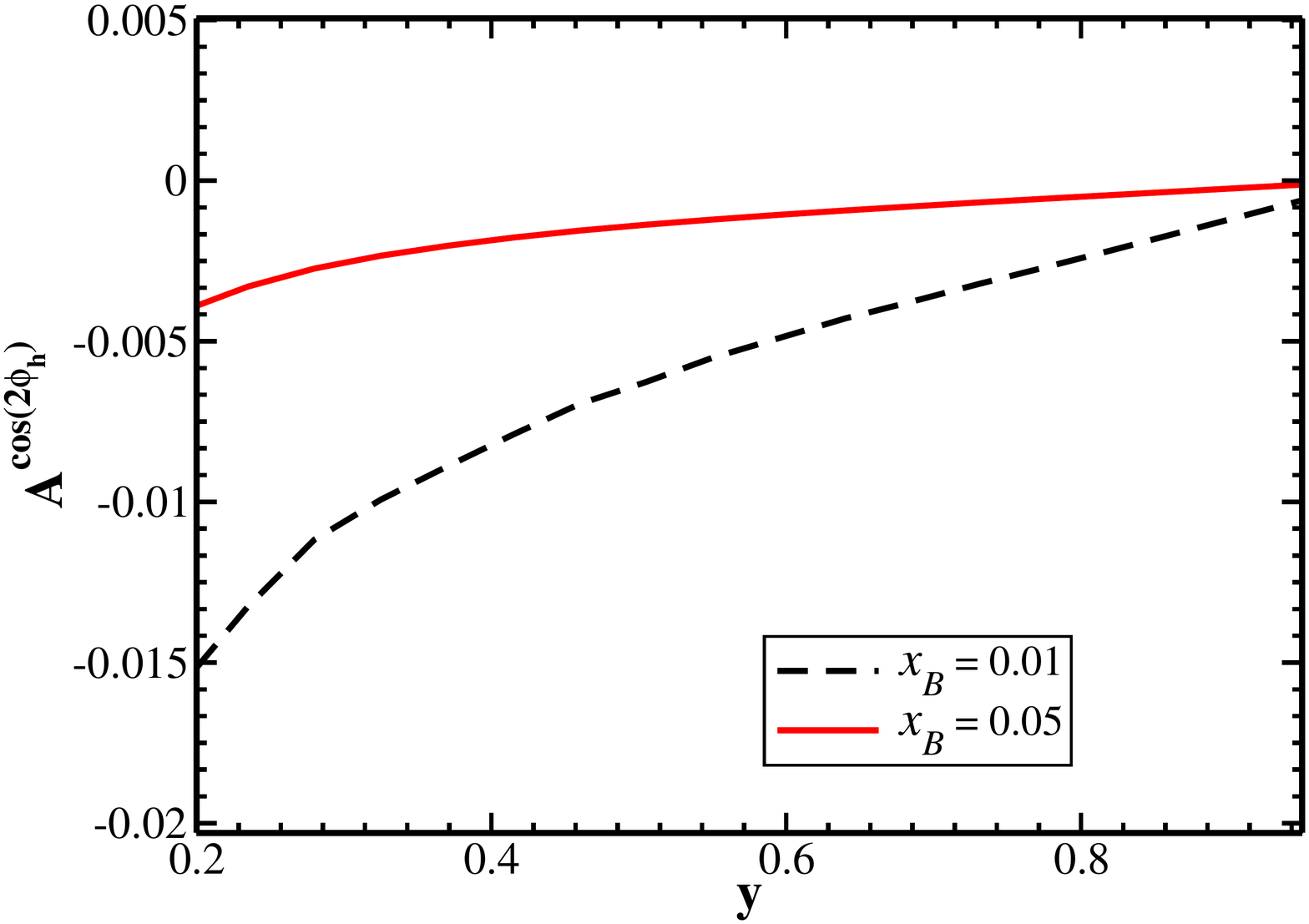}
  \caption{$cos(2\phi_h)$ asymmetry  in $e+p\rightarrow e+ J/\psi +X$
		process. Left: As a function of $P_{h\perp}$ at $\sqrt{s}=100$ GeV and $Q^2=15$ GeV$^2$.
 		The integration ranges are $0.0015 < x_B < 0.1$, $0.1 < z < 0.9$ and y is set by the values of $Q^2$, $s$ and $x_B$. Right: As a function of $y$ at $\sqrt{s}=120$  GeV and at fixed $x_B$ .
 		The integration ranges are $0 < P_{h\perp} < 2$, $0.1 < z < 0.9$. For both plots CMSWZ set of LDMEs \cite{chao2012j} is used.}
 \label{g1}

 \end{figure}
In Fig. \ref{g1}-\ref{g3}, we have used the Gaussian parameterization of the TMDs. In Fig. \ref{g1} and \ref{g3} we calculated the total asymmetry by including the contribution from both the color singlet and color octet states in the framework of NRQCD. \\
In left panel of Fig. \ref{g1} we showed the asymmetry as a function of $P_{h \perp}$ at center of mass energy $\sqrt{s}=100$ GeV and at a fixed value of $Q^2=15$ GeV$^2$.  The integration ranges are $0.0015 < x_B < 0.1$, $0.1 < z < 0.9$ and $y$ is set by the values of $Q^2$, $s$ and $x_B$. In the right panel of Fig. \ref{g1}, we showed the asymmetry as a function of $y$ at $\sqrt{s}=120~$ GeV and at fixed $x_B$. The asymmetry is larger for smaller values of $x_B$ and approaches to zero as $y$ tends to 1.  We obtained  negative asymmetry which is consistent with the LO calculations shown in \cite{mukherjee2017j}.

\begin{figure}[H]
 
 \centering
 \includegraphics[width=.50\textwidth]{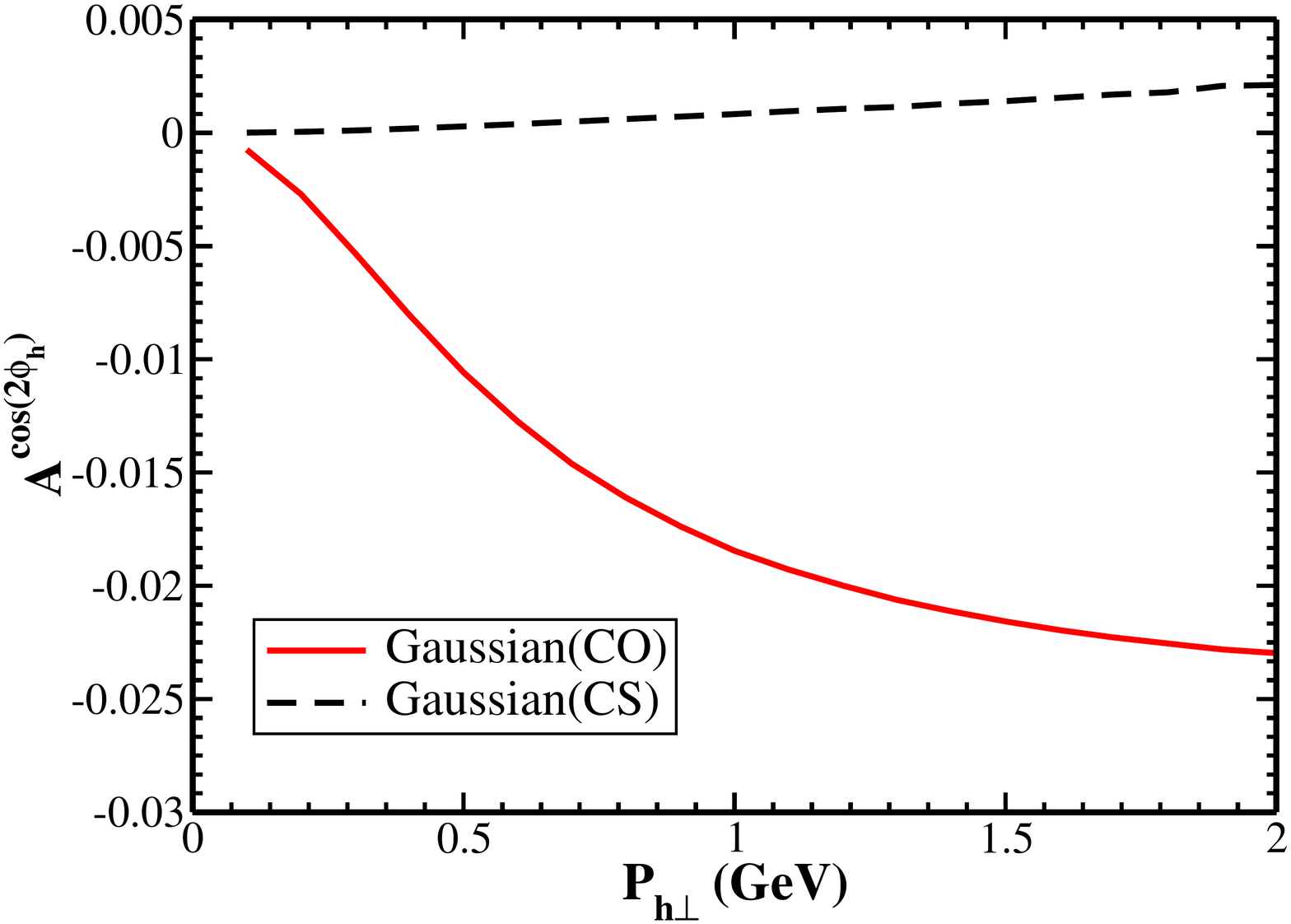}\hfill
 \includegraphics[width=.50\textwidth]{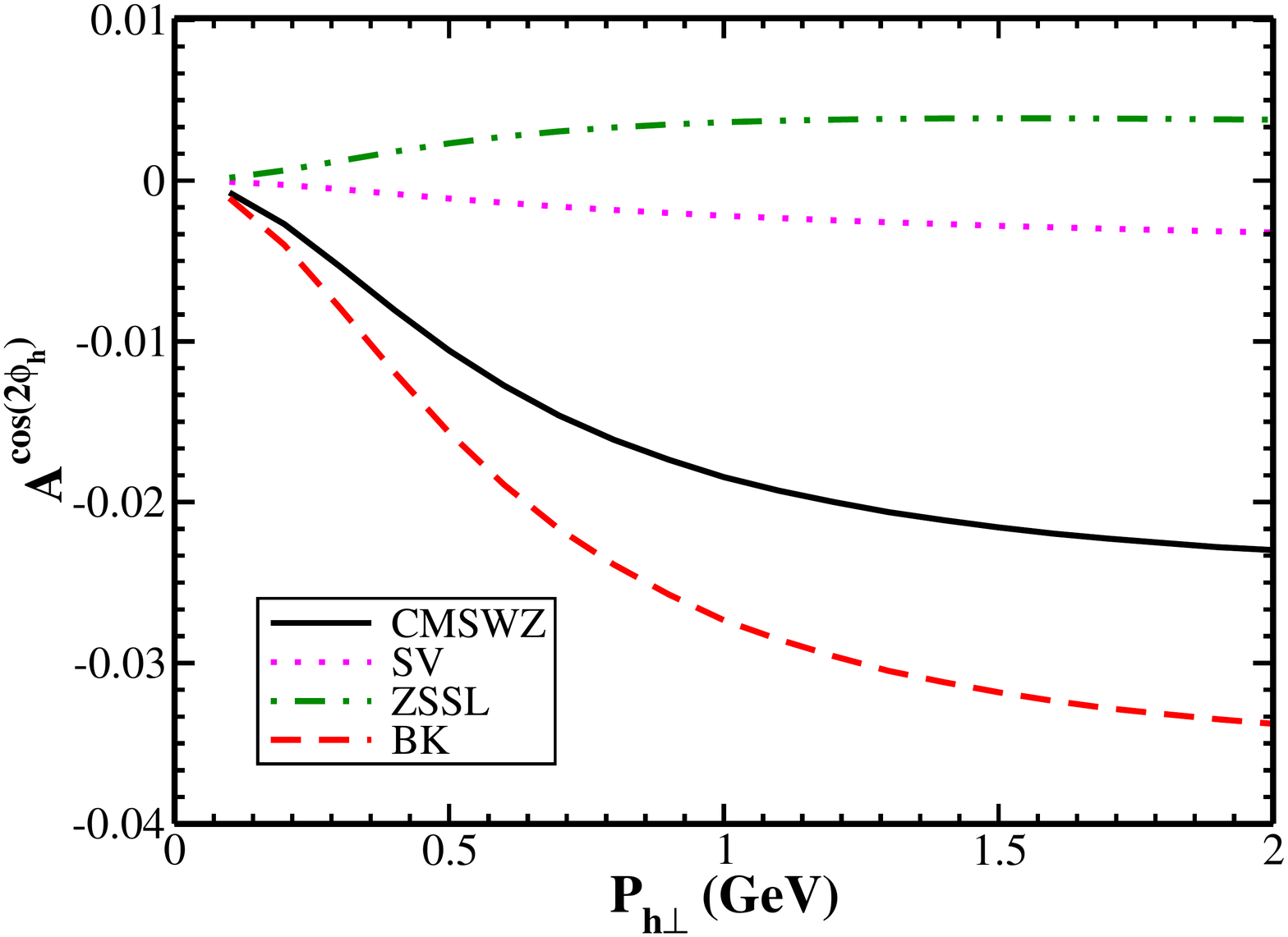}\hfill

 \caption{$cos(2\phi_h)$ asymmetry  in $e+p\rightarrow e+ J/\psi +X$
		process as function of $P_{h\perp}$ at $\sqrt{s}=100$ GeV and $Q^2=15$ GeV$^2$.
 		The integration ranges are $0.0015 < x_B < 0.1$, $0.1 < z < 0.9$ and y is set by the values of $Q^2$, $s$ and $x_B$.
  		Left:  contributions from color singlet and color octet states  using the LDMEs set CMSWZ \cite{chao2012j}. Right: comparing the asymmetry for  different LDMEs sets: CMSWZ \cite{chao2012j}, SV \cite{sharma2013high}, ZSSL \cite{zhang2015impact}, BK \cite{butenschoen2011world}.}
 \label{g2}
 \end{figure}
 In the left panel of Fig. \ref{g2}, we present the contribution to the asymmetry from  the color singlet(CS) and the color octet(CO) states. From the plot, we see that  the color octet states are  giving a significant contribution to the asymmetry, whereas the contribution from the CS is almost zero and slightly positive in higher $P_{h \perp}$ region. In the right panel of Fig. \ref{g2}, we showed the asymmetry for different sets of LDMEs. We see that the magnitude and the sign of the asymmetry depends on the set of LDMEs used. In fact, this is because of different states contributing to the asymmetry  depending on the LDMEs. We have a larger asymmetry for LDMEs set BK \cite{butenschoen2011world}. Also, in Ref. \cite{mukherjee2017j} and Ref. \cite{DAlesio:2019qpk}, it has been shown that at LO and at NLO, respectively, in the color octet model, the asymmetry comes from the several states and the results depend on the choice of LDMEs.

 \begin{figure}[H]
 
 \centering
 \includegraphics[width=.50\textwidth]{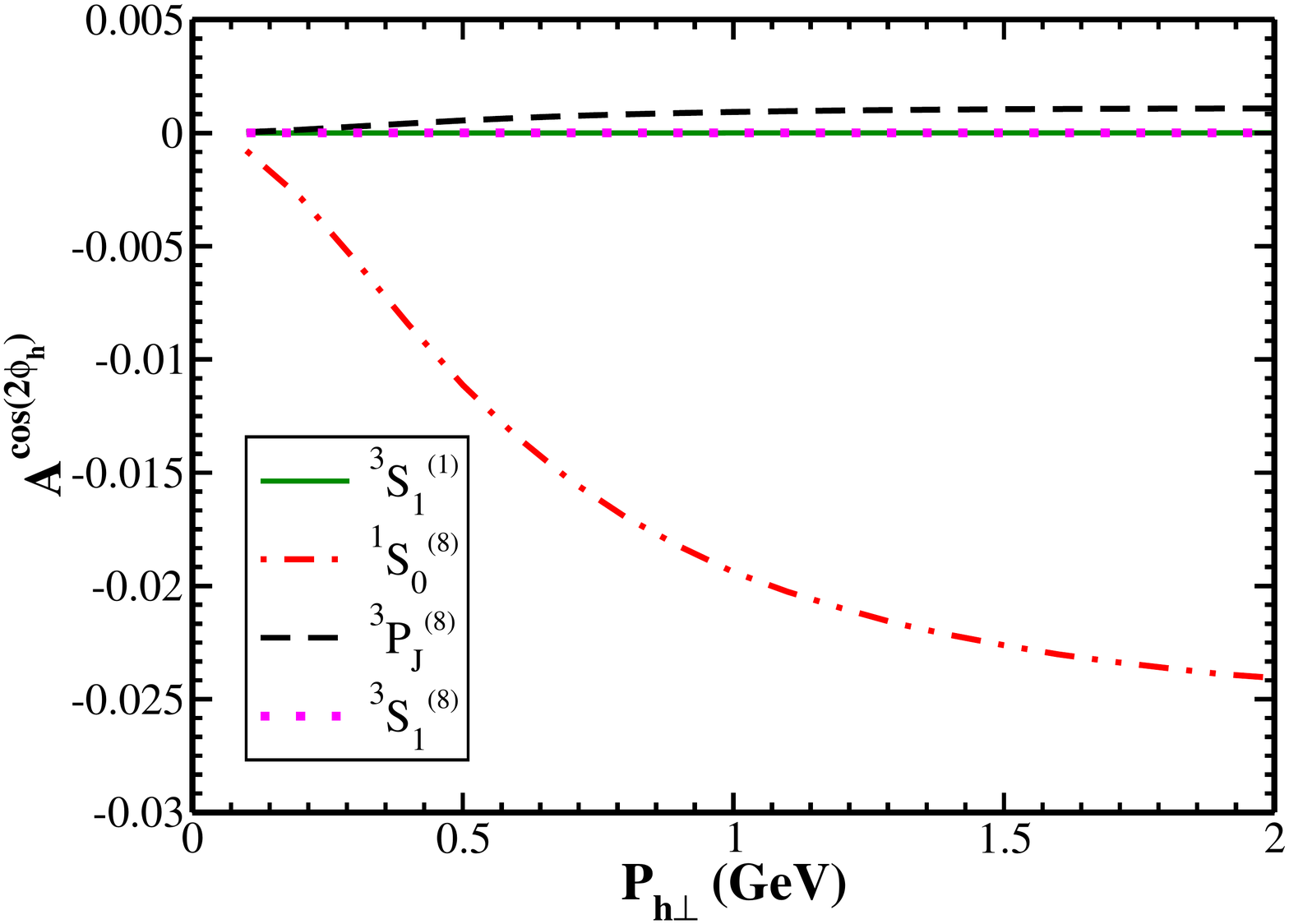}\hfill
    \includegraphics[width=.50\textwidth]{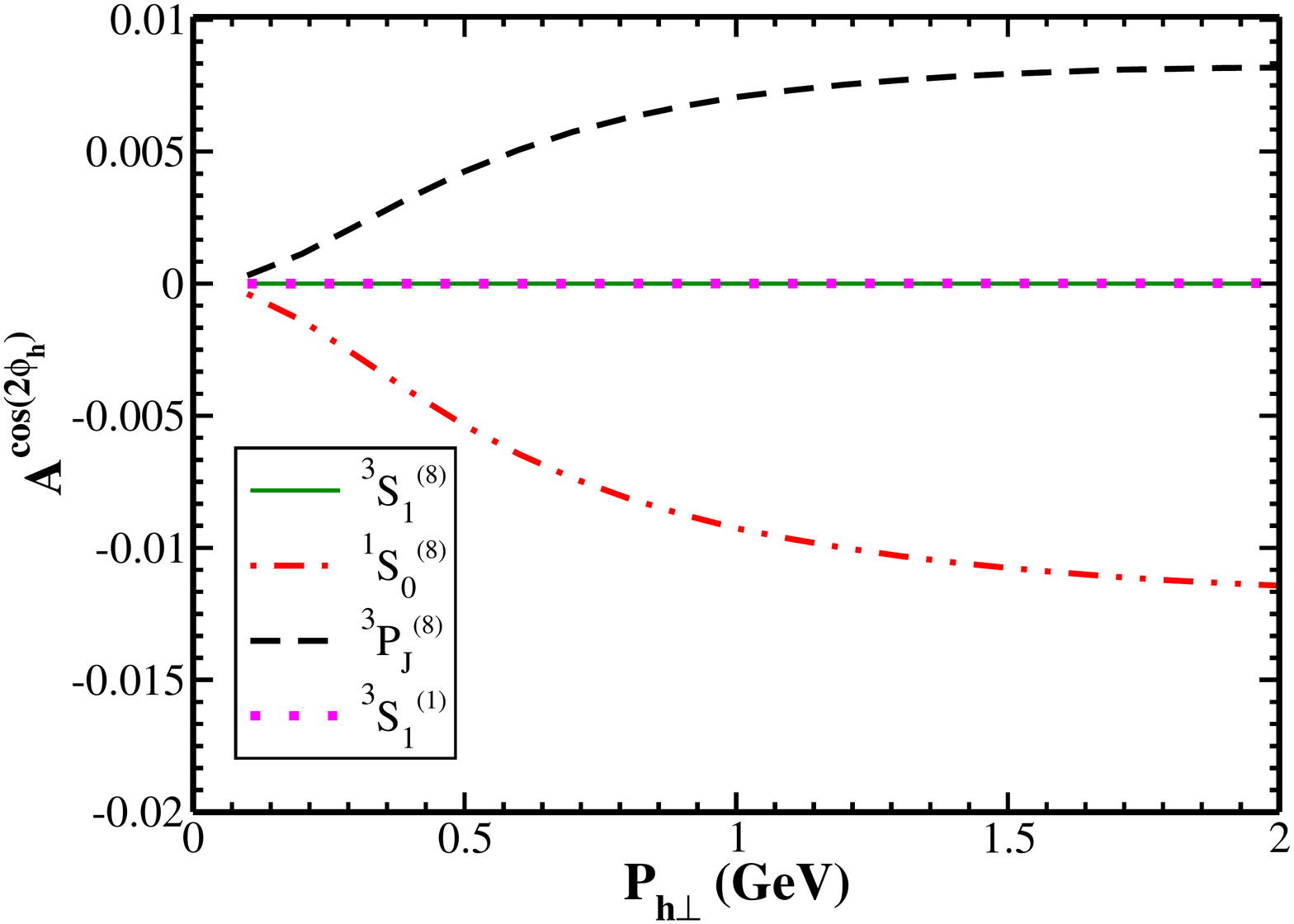}\hfill

 \caption{$cos(2\phi_h)$ asymmetry  in $e+p\rightarrow e+ J/\psi +X$
		process as function of $P_{h\perp}$ at $\sqrt{s}=100$ GeV and $Q^2=15$ GeV$^2$.
 		The integration ranges are $0.0015 < x_B < 0.1$, $0.1 < z < 0.9$ and y is set by the values of $Q^2$, $s$ and $x_B$. Left: asymmetry for the set of LDMEs, CMSWZ \cite{chao2012j}. Right: asymmetry for the set of LDMEs SV \cite{sharma2013high} }
 \label{g4}
 \end{figure}
In the left panel of Fig. \ref{g4}, we show the contribution coming from all the individual states to the asymmetry for the LDMEs set CMSWZ \cite{chao2012j}. For this set of LDMEs,  there is a dominance of one single state, $^1S_0^{(8)}$ to the asymmetry. 
Whereas for the LDMEs set SV \cite{sharma2013high}, which is shown in the right panel of the same figure, we have major contributions from two states $^1S_0^{(8)}$ and $^3P_J^{(8)}$. 
In Fig. \ref{g3}, we plotted the asymmetry as a function of $P_{h \perp}$ for different values of the parameter $r$(left) and  Gaussian width $\langle k_\perp^2\rangle$(right). We do not see any significant change in the asymmetry over different values of Gaussian width and the parameter $r$.\\
Also, we have checked that  there is no significant change in the asymmetry if the value of the centre of mass energy is changed for a fixed value of $Q^2$; but the asymmetry decreases  with the increase of the $Q^2$  for a fixed value of $s$.  

\subsubsection{$cos(2\phi_h)$ asymmetry in the McLerran-Venugopalan (MV) model}
\begin{figure}[H]
 
 \centering
 \includegraphics[width=.50\textwidth]{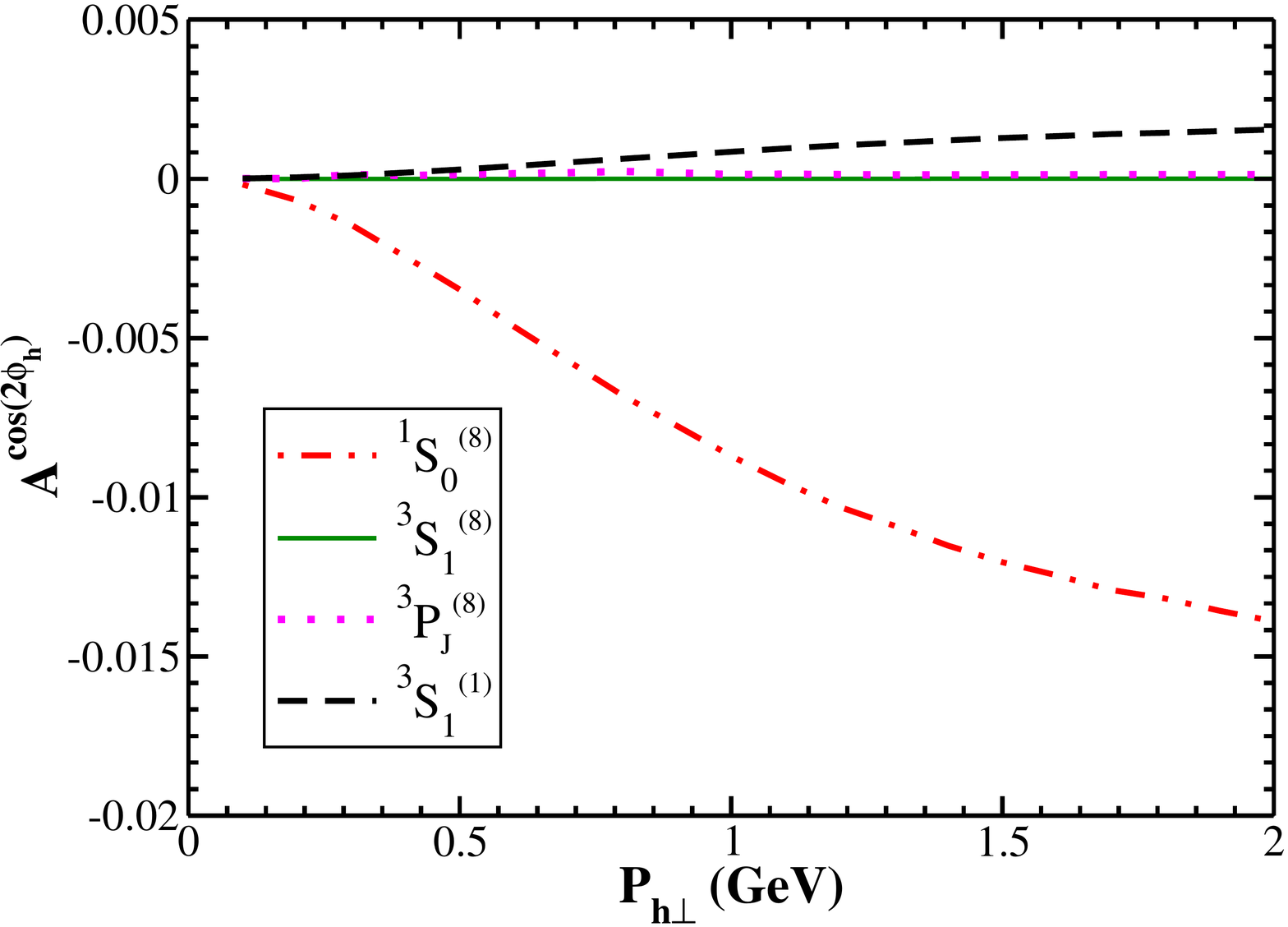}\hfill
   \includegraphics[width=.50\textwidth]{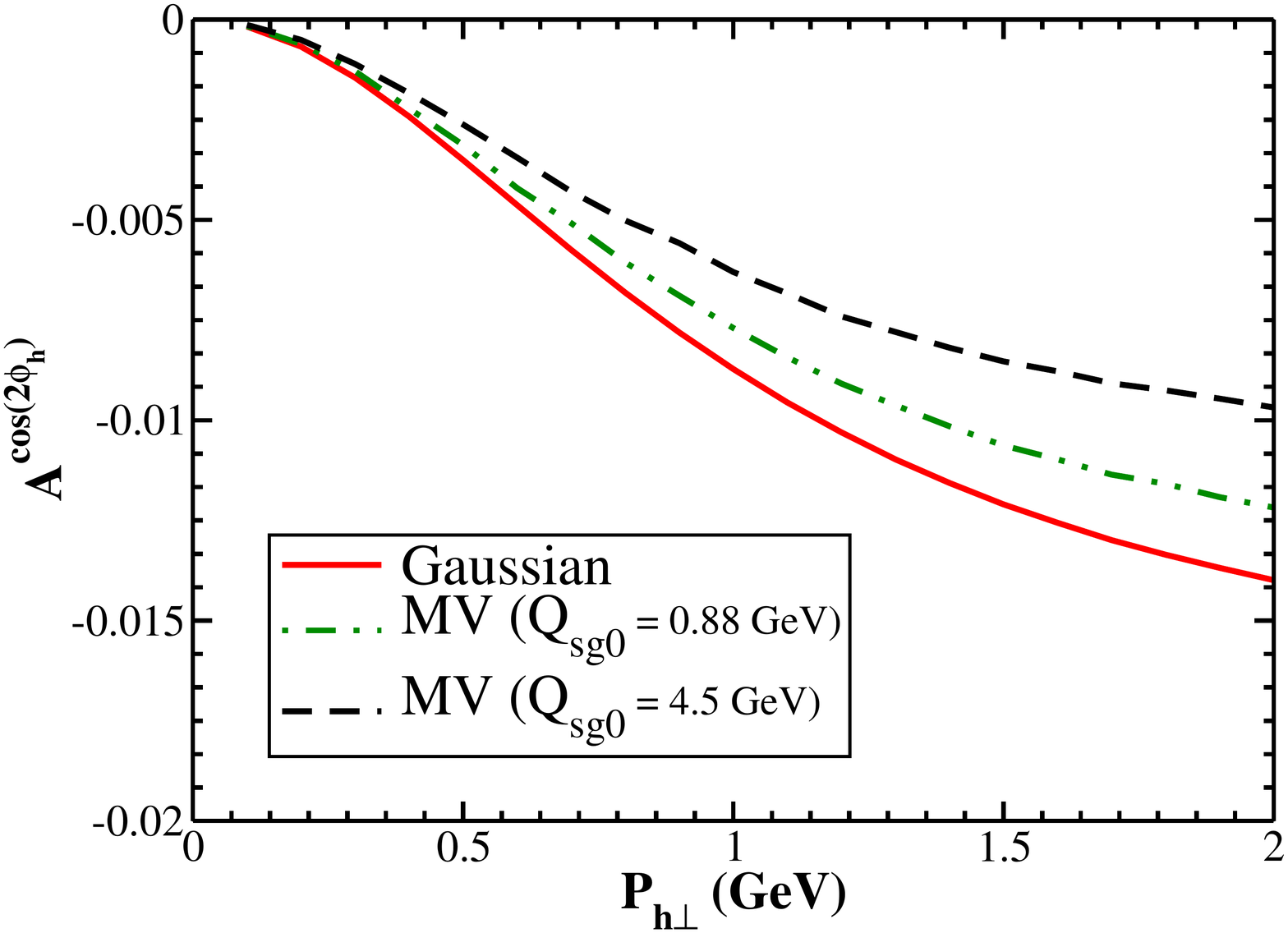}\hfill

 \caption{$cos(2\phi_h)$ asymmetry  in $e+p\rightarrow e+ J/\psi +X$
 		process as function of $P_{h\perp}$ at $\sqrt{s}=150$ GeV, $x=0.01$ and $z=0.7$. 
		Left: contribution to the $cos(2\phi)$ asymmetry coming from the individual states, as a function of $P_{h \perp}$ in the NRQCD framework using color octet model. 
        Right: comparison of the Gaussian and MV model (with two different values of $Q_{sg0}$). For both plots, CMSWZ set of LDMEs \cite{chao2012j} is used. }
 \label{mv2}

 \end{figure}

\begin{figure}[H]
 
 \centering
  \includegraphics[width=.5\textwidth]{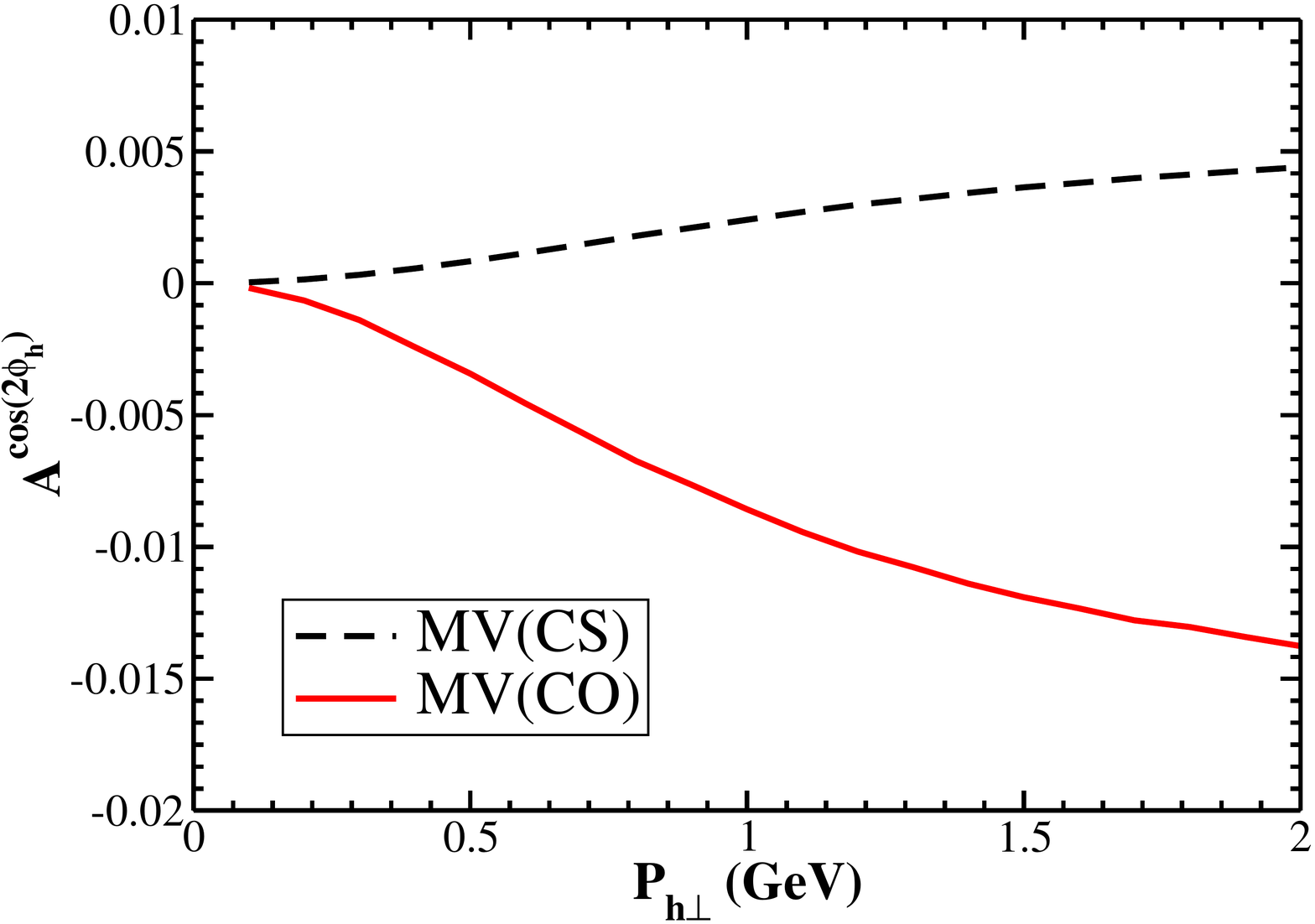}\hfill

 \caption{$cos(2\phi_h)$ asymmetry  in $e+p\rightarrow e+ J/\psi +X$
 		process as function of $P_{h\perp}$ at $\sqrt{s}=150$ GeV, $x=0.01$ and $z=0.7$. Contribution to the $cos(2\phi_h)$ asymmetry coming from color singlet and color octet model. CMSWZ set of LDMEs \cite{chao2012j} is used.}
 \label{mv3}
 \end{figure}

The McLerran-Venugopalan (MV) model \cite{mclerran1994computing,mclerran1994gluon,mclerran1994green}, a classical model, helps us to calculate the gluon distribution in a large nucleus at small $x$. In this model, the Gaussian distribution of color charges is assumed which act like a static sources, producing the soft gluons by the Yang-Mills equations. Although originally proposed for a large nucleus, this model has been found to give reasonable phenomenological results in azimuthal asymmetries for a proton in small-$x$ domain \cite{pisano2013linear,kishore2019accessing,DAlesio:2019qpk}.  The analytical expressions for unpolarized and linearly polarized Weizs$\ddot{a}$cker-Williams  gluon distribution can be written as, following  \cite{metz2011distribution,dominguez2012linearly} :
\begin{eqnarray}
 f_1^g(x,\bold{k}_\perp^{2})&=&\frac{S_{\perp}C_{F}}{\alpha_s\pi^3}\int d\rho\frac{J_0(k_{\perp}\rho )}{\rho}\left(1-\exp^{\frac{-\rho^2}{4}Q^2_{sg}(\rho)}\right)\\
 h_1^{\perp g}(x,\bold{k}_\perp^{2})&=&\frac{2S_{\perp}C_{F}}{\alpha_s\pi^3}\frac{M^2_P}{k_{\perp}^2}\int d\rho\frac{J_2(k_{\perp} \rho)}{\rho \log\left(\frac{1}{\rho^2\lambda^2_{QCD}}\right)}
 \left(1-\exp^{\frac{-\rho^2}{4}Q^2_{sg}(\rho)}\right)
\end{eqnarray}
here $S_{\perp}$ is the transverse size of the  proton, $Q_{sg}(\rho)$ is the saturation scale for gluons, which in general is a function of $x$ but in MV model depends on dipole size $\rho$ logarithmically. Here $Q^2_{sg}(\rho)=Q^2_{sg0}\ln\left(1/\rho^2\lambda^2+\epsilon\right)$, where $Q^2_{sg0}=(N_c/C_F)Q^2_{s0}$ with $Q^2_{s0}=0.35$ GeV$^2$ at $x=x_0=10^{-2}$ and $\lambda_{QCD}=0.2$ GeV from the HERA data \cite{golec1998saturation} fitting. Following the approach of \cite{DAlesio:2019qpk}, a regulator $\epsilon$ is added for numerical convergence. The results for MV parameterization are obtained in the kinematic region defined by $\sqrt{s}=150$ GeV, $x=0.01$ and $z=0.7$. The integration ranges are $y\in[0.2,0.9]$ and $x_B\in[0.005,0.009]$. The value of $Q$ is set according to  $y,x_B$ and $s$.  In Fig. \ref{mv2} (Left), we present the contribution to the $cos(2\phi_h)$ asymmetry for the $J/\psi$ coming from the individual states, as a function of $P_{h \perp}$. The maximum contribution to the asymmetry comes from the  $^1S^{(8)}_0$ state. In the same Fig. \ref{mv2} (right), we show the  comparison of  the asymmetry in the Gaussian and MV model, respectively (with two different values of $Q_{sg0}$\cite{golec1998saturation,albacete2005numerical}) within the same kinematical region as discussed above.  The asymmetry in MV model depends on the saturation scale. We note that the Gaussian parameterization of the TMDs gives larger   $cos(2\phi_h)$ asymmetry. In Fig. \ref{mv3} we compared the contribution to the asymmetry from  color singlet and color octet states. We found that larger contribution to the $cos(2\phi_h)$ asymmetry comes from color octet states, which is negative. Color singlet states give a small positive contribution to the asymmetry. 

\section{conclusion}
In this article, we studied  the $cos(2\phi_h)$  asymmetry in $J/\psi$ production in an electron-proton collision in the kinematics of the future electron-ion collider. We calculated them in the small-$x$ domain, where the gluon TMD, namely the unpolarized and linearly polarized gluon TMD are important in unpolarized scattering.  The dominant sub-process in this kinematical region for $J/\psi$ production is the virtual-photon-gluon fusion process $\gamma^{\ast}+g\rightarrow J/\psi+g$. 
We used the NRQCD  based color octet formalism  for calculating the $J/\psi$ production rate. We obtain a small but sizable
$cos(2\phi_h)$ asymmetry. The asymmetry depends on the parameterization of the gluon TMDs used. We used both the Gaussian as well as the MV model for the parameterization. The magnitude of the asymmetry was found to be larger for Gaussian parameterization. We included contributions both from CO as well as CS states.  The asymmetry depends on the LDMEs used, in particular, contributions from individual states were found to depend substantially on the set of LDMEs used. Overall, our calculation shows that the $cos(2 \phi_h)$ asymmetry in $J/\psi$ production could be a very useful tool to probe the ratio of the linearly polarized gluon TMD and the unpolarized gluon TMD  in the small-$x$ region at the EIC. 

\bibliographystyle{apsrev}
\bibliography{biblography_as.bib}

\end{document}